\documentclass[aps,pre, reprint, showpacs,superscriptaddress, nofootinbib,floatfix]{revtex4-1} 

\usepackage{latexsym}
\usepackage{amsmath,amsfonts,amssymb,amsbsy,amscd}
\usepackage{hyperref}
\usepackage{graphicx}
\usepackage{gensymb}
\usepackage{color}
\hypersetup{
   pdfauthor=Siminos,
   pdftitle=Modeling ultrafast shadowgraphy
}

\definecolor{hreflinkcolor}{rgb}{0.13,0.17,0.83}
\hypersetup{colorlinks=true,urlcolor=hreflinkcolor,
		linkcolor=hreflinkcolor,citecolor=hreflinkcolor}

\newcommand{\beq}{\begin{equation}}
\newcommand{\eeq}{\end{equation}}
\newcommand{\bseq}{\begin{subequations}}
\newcommand{\eseq}{\end{subequations}}

\newcommand{\rf}     [1] {~\cite{#1}}
\newcommand{\refref} [1] {Ref.~\cite{#1}}

\newcommand{\refeq}  [1] {Eq.~(\ref{#1})}

\newcommand{\refeqs} [2]{Eqs.~(\ref{#1})--(\ref{#2})}

\newcommand{\reffig} [1] {Fig.~\ref{#1}}

\newcommand{\refFig} [1] {Figure~\ref{#1}}

\newcommand{\reftab} [1] {Table~\ref{#1}}

\newcommand{\refsect}[1] {Sec.~\ref{#1}}

\newcommand{\ie}{{i.e.}}        %
\newcommand{\cf}{{\em cf.\ }}     %
\newcommand{\eg}{{e.g.}}        %

\newcommand{\micron}{\ensuremath{\mathrm{\mu m}}}

\newcommand{\tpr}{\ensuremath{\tau_{\mathrm{pr}}}}
\newcommand{\lambdaPr}{\ensuremath{\lambda_{\mathrm{pr}}}}
\newcommand{\fs}{\ensuremath{\mathrm{fs}}}

\begin{document}

\author{E.~Siminos}
\email{evangelos.siminos@gmail.com}
\affiliation{Max Planck Institute for the Physics of Complex Systems, N\"{o}thnitzer Str. 38,
D-01187 Dresden, Germany}

\author{S.~Skupin}
\affiliation{Univ.~Bordeaux - CNRS - CEA, Centre Lasers Intenses et Applications, UMR 5107, 33405 Talence, France}

\author{A.~S\"{a}vert}
\affiliation{Institut f\"{u}r Optik und Quantenelektronik, Abbe-Center of Photonics, Friedrich-Schiller-Universit\"{a}t, 07743 Jena, Germany}

\author{J.~M.~Cole}
\affiliation{The John Adams Institute for Accelerator Science, The Blackett Laboratory, Imperial College London, London SW7 2AZ, United Kingdom}

\author{S.~P.~D.~Mangles}
\affiliation{The John Adams Institute for Accelerator Science, The Blackett Laboratory, Imperial College London, London SW7 2AZ, United Kingdom}

\author{M.~C.~Kaluza}
\affiliation{Institut f\"{u}r Optik und Quantenelektronik, Abbe-Center of Photonics, Friedrich-Schiller-Universit\"{a}t, 07743 Jena, Germany}
\affiliation{Helmholtz-Institut Jena, Friedrich-Schiller-Universit\"{a}t, 07743 Jena, Germany}

\title{Modeling ultrafast shadowgraphy in laser-plasma interaction experiments}

\begin{abstract}
Ultrafast shadowgraphy is a new experimental technique that uses few cycle laser pulses to image density gradients in a rapidly evolving plasma. It enables structures that move at speeds close to the speed of light, such as laser driven wakes, to be visualized. Here we study the process of shadowgraphic image formation during the propagation of a few cycle probe pulse transversely through a laser-driven wake using three-dimensional particle-in-cell simulations. In order to construct synthetic shadowgrams a near-field snapshot of the ultrashort probe pulse is analyzed by means of Fourier optics, taking into account the effect of a typical imaging setup. By comparing synthetic and experimental shadowgrams we show that the generation of synthetic data is crucial for the correct interpretation of experiments. Moreover, we study the dependence of synthetic shadowgrams on various parameters such as the imaging system aperture, the position of the object plane and the probe pulse delay, duration and wavelength. Finally, we show that time-dependent information from the interaction can be recovered from a single shot by using a broadband, chirped probe pulse and subsequent spectral filtering.
\end{abstract}

\pacs{42.30.Kq, 42.30.Va, 52.38.Kd, 41.75.Jv, 52.65.Rr, 52.70.-m}

\date{\today}

\maketitle

\section{Introduction}

The interaction of intense laser pulses with plasma can lead to the excitation of nonlinear structures,
such as wakes, solitons, shocks and filaments~\cite{mourou2006}. The time evolution of such structures is highly non-trivial
and a lot of effort has been invested in developing theoretical tools and experimental diagnostics to investigate it.
When such nonlinear structures move close to the speed of light $c$, as is the case for a wake driven by an intense
laser pulse in an underdense plasma, additional difficulties arise due to the requirement of short interaction times and micrometer length-scales in 
order to resolve the relevant dynamics.

In a typical laser-wakefield acceleration (LWFA) experiment, a relativistically intense laser pulse (the `pump' pulse) excites an electron
plasma wave (the `wake') which propagates with a phase velocity $v_p$ equal to 
the pump pulse's group velocity $v_g\simeq c$ through the plasma. Ambient electrons can be trapped
and accelerated in the strong electromagnetic fields of the wake, producing quasi-monoenergetic beams~\rf{mangles2004,geddes2004,faure2004} 
up to an energy of $4\,\mathrm{GeV}$~\rf{leemans2014}.
Stable operation of such accelerators requires a thorough understanding of nonlinear laser-plasma interaction processes 
motivating the development of novel diagnostics. 
The characteristic length scale of the wake is the plasma wavelength $\lambda_p$. 
Therefore, optical probing of such a wake with a transversely propagating probe pulse requires 
a probe pulse duration $\tpr\ll\lambda_p/c$. 
Typically this corresponds to a duration $\tpr$ of a few optical cycles $\tau_L=2\pi/\omega_L$, where $\omega_L$ is the
center frequency of the probe pulse. Recent pump-probe experiments utilized probe pulses of few femtosecond 
duration to obtain the first direct shadowgraphic images 
of a laser-driven wake propagating in an underdense plasma~\cite{schwab2013,saevert2015}.
These measurements offer far more information about the plasma wave than previous ones
using longitudinal\rf{marques96,siders96,matlis06} or longer transverse probe pulses\rf{kaluza2010,buck11}.

Conventional shadowgraphy can be analyzed using geometrical optics and be formulated as
an inverse problem: for a given shadowgraphic image local gradients in index of refraction can be
retrieved, see, \eg, \refref{panigrahi2012}. In ultrafast shadowgraphy of laser induced wakes such retrieval is in general not straightforward.
The effect of longitudinal motion of the wake as the probe pulse traverses, the presence of strong magnetic fields
as well as relativistic effects are not negligible
and need to be taken into account. At the same time diffraction effects can be significant
when the probe pulse wavelength \lambdaPr\ is comparable to the characteristic length scale 
of density gradients in the plasma, \ie, the skin depth $\lambda_s=c/\omega_{pe}$, 
where $\omega_{pe}=\left(q_e^2 n_0/\epsilon_0 m_e\right)^{1/2}$
is the electron plasma frequency for a plasma of density $n_0$.
For parameters used in recent experiments~\cite{saevert2015} this is the case
and the effect of diffraction has to be accounted for.

In this paper we analyse how ultra fast probe pulses form shadowgraphic images by using 3D particle-in-cell (PIC)
simulations of the full pump-probe interaction.
Propagation of the pump pulse through the plasma is fully simulated and at different delays
a probe pulse propagating transversely to the pump is launched. 
It is worth noticing that the computational overhead due to the probe pulse is very small, because only (transverse)
propagation through the interaction zone (plasma wake) needs to be simulated by the PIC code.
After the probe traverses
the wake, post-processing in Fourier space allows the
effect of a typical imaging setup in shadowgraphic image formation to be taken into account.
We show that synthetic shadowgrams generated using this methodology 
turn out to be crucial for the correct interpretation of experimental measurements.

This paper is structured as follows. In \refsect{s:simulations} we present our PIC simulations
of pump and probe propagation. The methodology we use to post-process the results
in order to obtain artificial shadowgrams is explained in \refsect{s:post}. 
The direct confrontation with experimental shadowgrams in 
\refsect{s:example} illustrates the importance of synthetic shadowgrams in the interpretation of experimental results. 
In \refsect{s:probeChar} we present several
case studies illustrating the effect of varying object plane position, duration and wavelength 
of probe pulse on the shadowgrams and show that time-dependent information can be recovered
from a single chirped probe pulse by spectral filtering.
Finally, in \refsect{s:concl} we discuss our 
findings and present our conclusions.

\section{\label{s:simulations}PIC simulations}

\subsection{Pump propagation}

We perform simulations of LWFA using the PIC code EPOCH~\cite{epoch} with parameters similar to the ones
used in recent experiments that utilized ultrafast shadowgraphy to study injection in LWFA~\cite{saevert2015}.
We consider a plasma with super-Gaussian electron density profile $n_e= n_0 e^{-(x-x_0)^4/w_p^4}$ in propagation direction of the pump pulse, where
$n_0=1.7\times10^{19}\,\mathrm{cm}^{-3}$, $x_0=1.17\,\mathrm{mm}$ and $w_p=1.198\,\mathrm{mm}$ are taken from
a fit of a typical experimental density profile of \refref{saevert2015}.
The computational domain is a `sliding window' of size $150\times70\times70\,\mathrm{\mu m}^3$ moving at $c$. 
We use $2700\times525\times525$ cells and two particles per cell, with fifth order particle weighting to reduce noise.
Ions are considered as a stationary backgound.
A Gaussian pump laser pulse propagating along the $x$ direction, linearly polarized along the $y$ direction,
with an intensity full width at half maximum (FWHM) duration of $36\,\mathrm{fs}$ and a central wavelength $\lambda_L=810\, \mathrm{nm}$
is focused to a spot size of $18.84\,\mathrm{\mu m}$ (intensity FWHM) at $x_f=300\,\micron$.
The pump pulse's maximum intensity (in vacuum) is $I_0=2.5\times10^{18}\,\mathrm{W/cm^2}$.
To ensure better dispersion properties of both the pump and probe pulses, while keeping the resolution requirements
low, the sixth order (in space) finite-difference-time-domain (FDTD)
Maxwell solver available in EPOCH is used.

\subsection{\label{s:probe} Probe propagation}

\begin{figure}[ht!]
    \begin{center}
    \includegraphics[width=0.25\textwidth, clip=true]{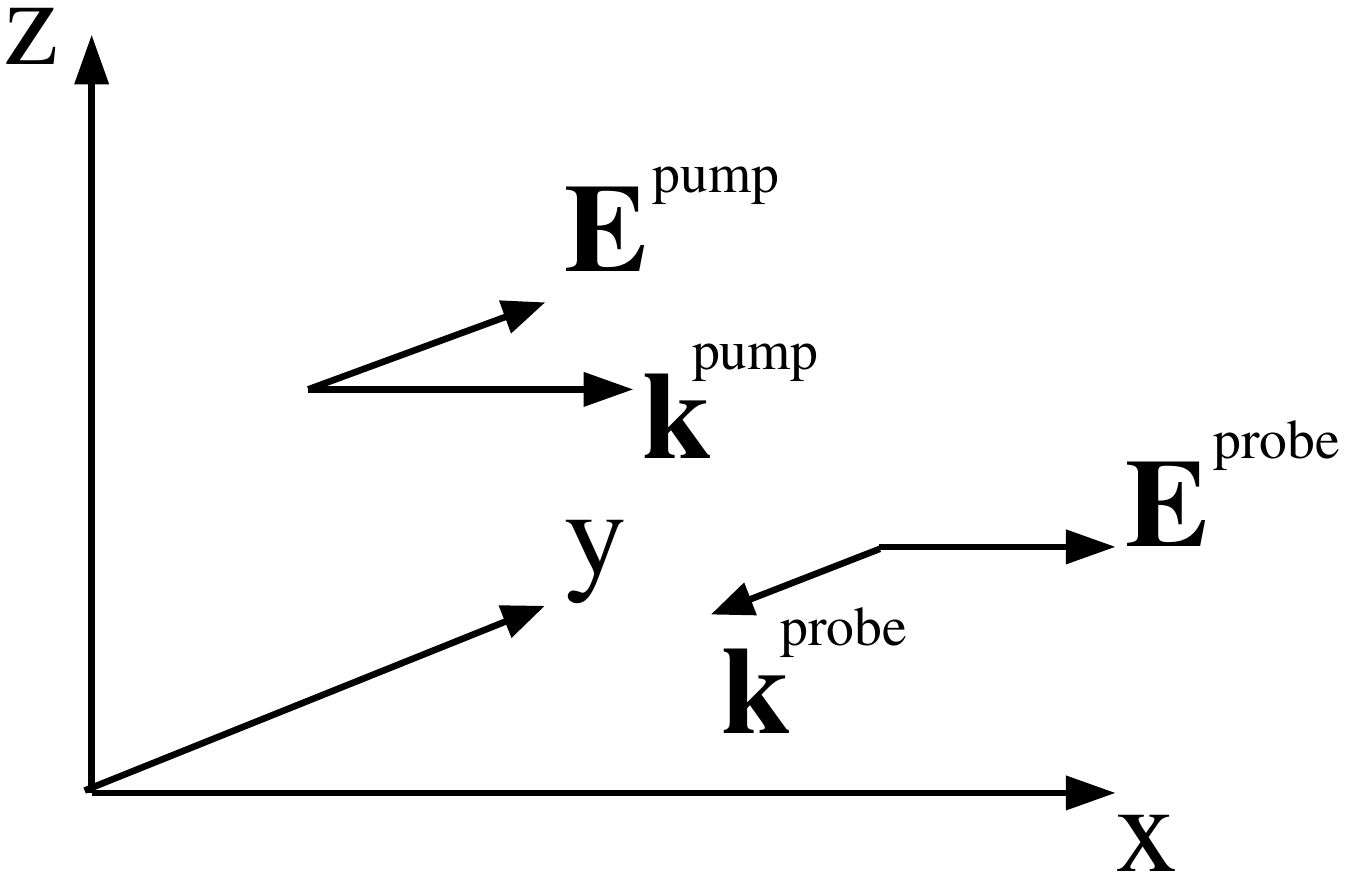}
  \end{center}
 \caption{\label{f:probeGeom}PIC simulation setup: The pump propagates along the $x$-direction and is polarized along $y$,
while the probe propagates along the (negative) $y$-direction and is polarized along~$x$.}
\end{figure}

Probe propagation in the plasma is also fully simulated in 3D with EPOCH. 
While simulating pump propagation, so called `restart dump' files, which 
contain the complete data representing the present state of the PIC simulation 
are saved at regular intervals. This allows us to start several probe simulations
at different `delays' using a single simulation of pump propagation.
For each probe simulation the moving window is stopped and a probe pulse
is injected from the side of the box, propagating along the negative $y$-direction, i.e. perpendicularly to the pump propagation direction,
see \reffig{f:probeGeom}. 
Because the pump propagates 
during the probing stage (without a moving window) 
additional space along the $x$-direction is required. Also, depending on the 
probe pulse duration $\tpr$, enlarging the computational box in $y$-direction may be necessary.
Moreover, it is important to take into account diffraction of the probe pulse from the computational boundaries
along the $x$ and $z$ directions. To keep it to a minimum we use a flat-top spatial profile of the probe pulse 
with sinusoidal ramps of $10\,\micron$ length. In our reconstruction of the shadowgrams we only take into account the 
part of the computational domain that is not affected by numerical diffraction from the boundaries.
All this results in a larger
computational domain compared to a typical LWFA simulation for the same parameters, however, 
this larger domain is only necessary during the probe simulations.
Moreover, each probe simulation is still much smaller than the full pump simulation because the pump
propagates for millimetres whereas the probe only propagates for tens of microns.

Although we investigate the influence of various probe pulse parameters, see \refsect{s:probeChar}, in all other runs the 
pulse parameters are similar to the ones in recent experiments~\cite{saevert2015}: 
central wavelength $\lambdaPr=0.75\,\micron$, bandwidth limited duration of $\tau_{BL}=4.4\,\mathrm{fs}$ (intensity FWHM),
a negative linear chirp that increased its duration to $\tpr=12\,\mathrm{fs}$, and maximum intensity $I_{\mathrm{pr}}=8.6\times 10^{14}\,\mathrm{W/cm^2}$.
The transverse size of the plasma is limited to few tens of microns in both the experiments and our simulations,
and for such short propagation distances the use of sixth order (FDTD) solver guarantees probe propagation without significant
lengthening of the pulse due to numerical dispersion for relatively low spatial resolution (see, \eg, \refref{shlager2004} for a study of dispersion
characteristics of high order FDTD solvers). 
We note that our probe pulse intensity is four orders of magnitude higher than typical experimental values\rf{saevert2015}
in order to have a better signal-to-noise ratio in the simulated shadowgrams (in PIC simulations the noise level in the electromagnetic fields
is much higher than in experiments due to the discretization by macroparticles).
However, we verified that varying the probe intensity by one order of magnitude does not alter the shadowgrams, \ie, the results can 
be considered linear in the probe intensity (as one would indeed expect for non-relativistic probe pulses). We note that 
one has to be careful when using increased probe intensity if ionization effects are included in the simulations (neglected in the present study).

\begin{figure}[ht!]
    \begin{center}
    \includegraphics[width=\columnwidth, clip=true]{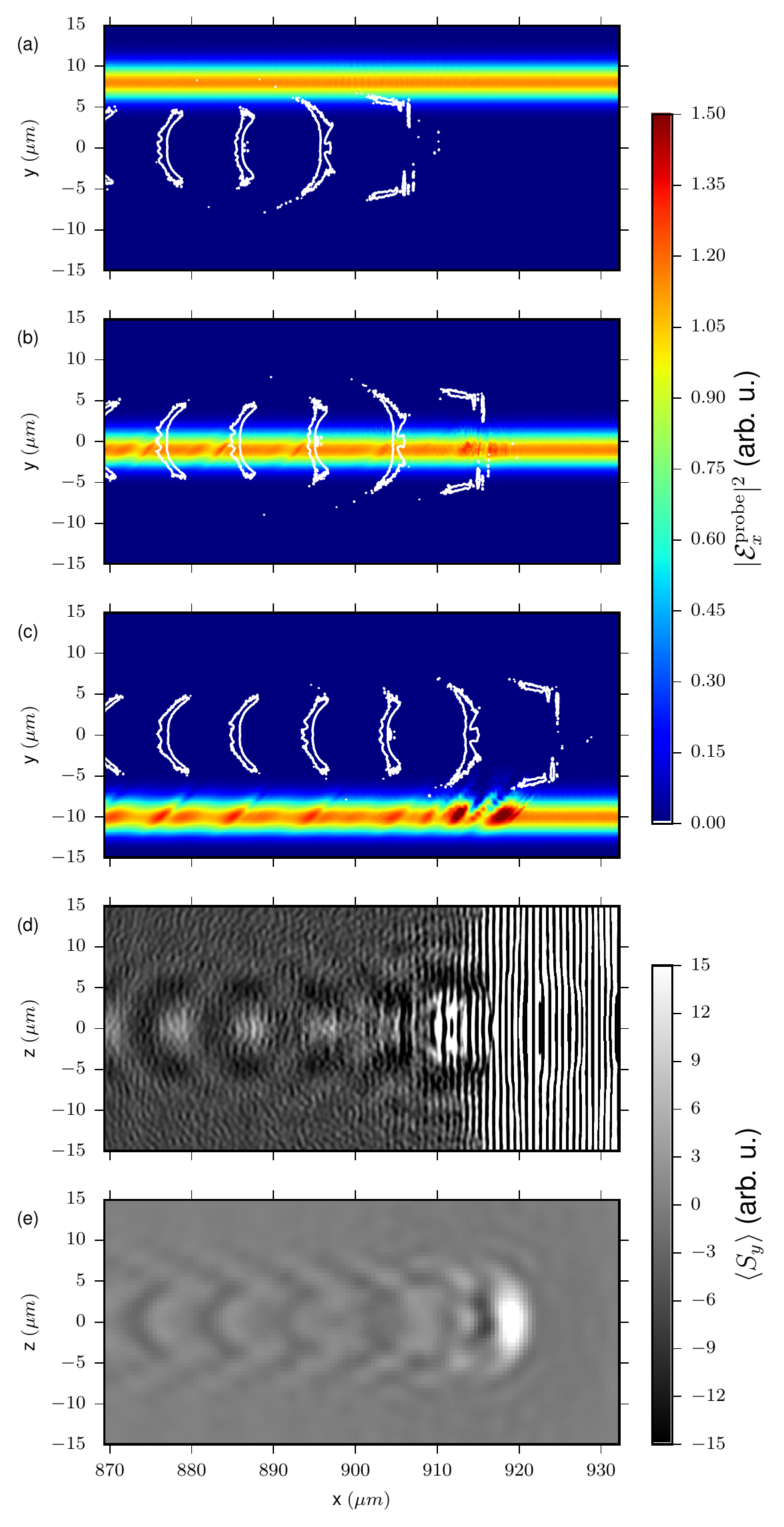} 
  \end{center}
 \caption{\label{f:probeTracking}Probe pulse propagation across the wake. Panels (a)-(c): Three snapshots from the PIC simulation showing
the probe electric field envelope squared $|\mathcal{E}_x^{\mathrm{probe}}|^2$ and contours of the plasma density corresponding to
$n=2\,n_0$. (d) Image obtained by recording the time-integrated Poynting flux $\langle{S}_y\rangle$ passing trough the plane $y_S=-20\,\micron$
(e) Time-integrated Poynting flux $\langle{S}_y\rangle$ after adjusting for focusing optics, assuming the object plane is at $y_o=0$.}
\end{figure}

In the simulations we allow the probe pulse to propagate past the wake structure, here approximately until its center reaches $y_S=-20\micron$. 
In \reffig{f:probeTracking}(a-c) we track the evolution of the envelope of the probe electric field (absolute of the complex electric field, see Sec.~\ref{s:method}) as it crosses the wake.
For presentation purposes only, we subtract the background fields using a simulation with identical initial conditions but no probe pulse.
\refFig{f:probeTracking}(b-c) indicates that modulations in the envelope of the probe pulse occur as it interacts with 
density gradients in the wake. The highest envelope modulations occur at the front of the bubble where the pump drives the wake.

Once the probe pulse reaches $y_S$ there are no
further density perturbations in the plasma, and therefore all local phase differences induced
by the wake have been imprinted to the probe pulse. 
However, if we simply 
try to reconstruct a shadowgram by recording the time-integrated intensity passing through the plane
$y_S$, as shown \reffig{f:probeTracking}(d), there are two problems with the image we obtain.
In the front of the wake there is strong scattering of pump light which is of much higher intensity 
than the probe intensity. This scattered light is not present in the experimental shadowgrams~\rf{schwab2013,saevert2015}.
As we will see the reason for this is that the aperture of the imaging system
eliminates most of the side-scattered light. Moreover, the wake structure appears blurred in
\reffig{f:probeTracking}(d). The reason for this is strong
diffraction, since the length scale for density gradients ($\lambda_s=1.3\,\micron$ here) is comparable to the
wavelength of the probe pulse ($\lambdaPr=0.75\micron$). 
In the experiments diffraction is compensated by optical imaging and
thus, in order to be able to compare our PIC simulation 
results to experimental shadowgrams, we have to take into account the
role of a typical imaging system in shadowgram formation. This could be done by 
solving a vacuum diffraction integration through the entire optical system. However, 
as we will describe in \refsect{s:post}, connecting the near field
results of our PIC simulations to the Fourier description of optical systems
offers a more straightforward and easy-to-implement method.

\section{\label{s:post} Post-processing and reconstruction of shadowgrams}

In order to reconstruct shadowgrams from our PIC simulation results, we use two major simplifications. 
Firstly, the transverse size of the simulation box in our PIC simulations is limited to a few of tens of microns, 
and so the PIC simulation cannot directly account for the optical imaging system. However, this is not really necessary, 
because once the probe pulse has traversed the plasma wake, all information related to the laser-plasma interaction process are already imprinted in its wavefronts.
At the same time, the wavefronts are distorted by diffraction, but this will be compensated by the imaging system. 
In order to take this into account,
we assume that the probe pulse propagates further on in vacuum, and neglect the influence of any low density plasma and/or 
gas which may be present between the main interaction region and the imaging system. By doing so, we can use much more efficient 
spectral pulse propagation methods. Secondly, the probe pulse, after it has passed the plasma wake, propagates mainly in one direction, 
and the imaging system has a certain numerical aperture (NA), i.e., acceptance angle. 
Here, we assume that the NA of our imaging system is small enough to justify a paraxial description of the imaging process.

\subsection{\label{s:method} Post-processing: Theory}

\begin{figure}[ht!]
    \begin{center}
    \includegraphics[width=\columnwidth, clip=true]{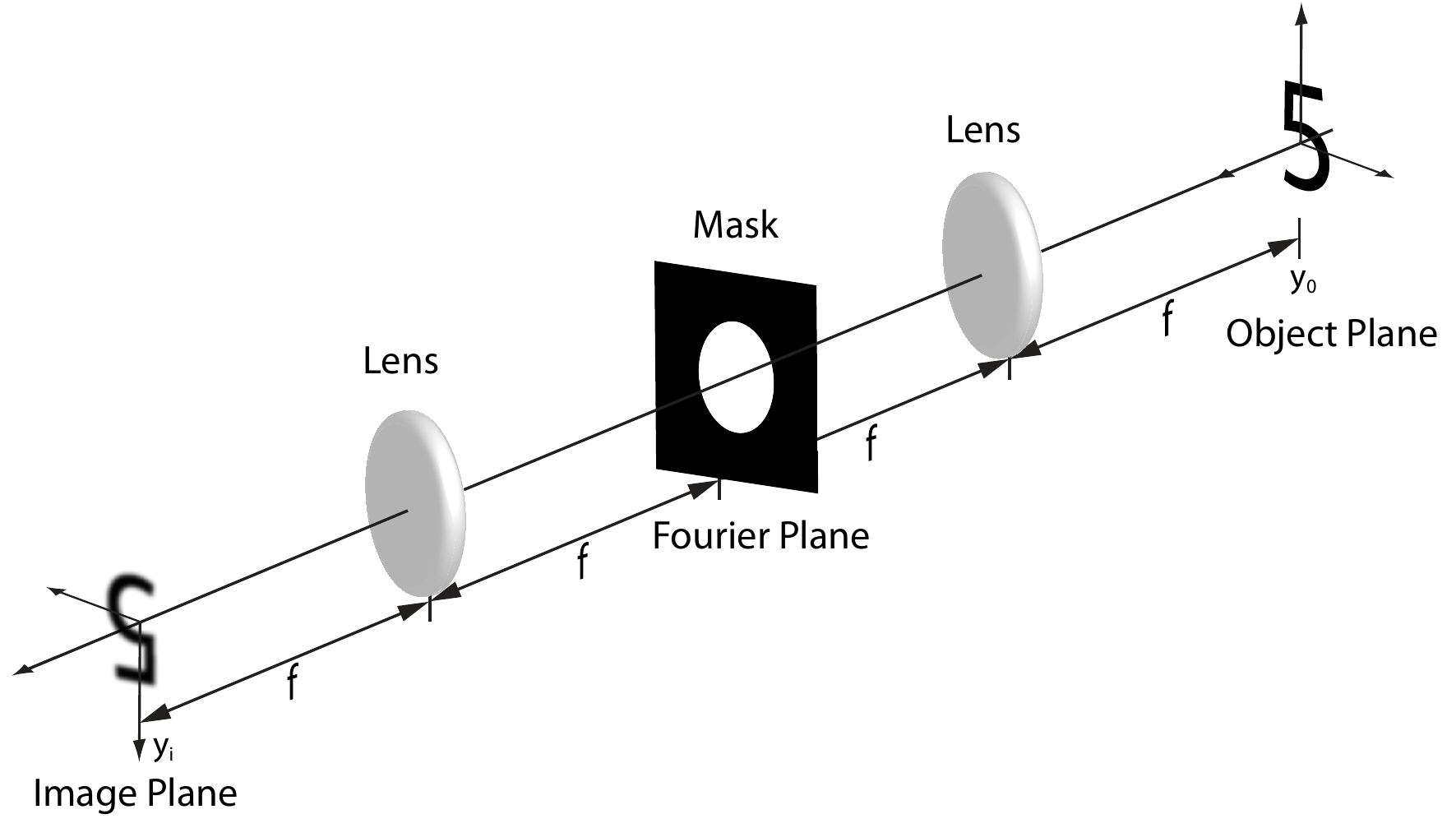} 
  \end{center}
 \caption{\label{f:4fsystem}Schematic setup of a so-called $4f$-system. In paraxial approximation 
 and by using thin-element approximation for the two lenses, we find the spatial Fourier transform 
 of the object plane in the Fourier plane, and a (mirrored) image in the image plane. 
 The aperture of the system can be represented by a (thin) mask in the Fourier plane. 
 }
\end{figure}

An optical imaging system in the paraxial approximation can be conveniently described 
by means of Fourier optics. 
Here, we will restrict ourselves to a generic $4f$ imaging system, which can be found in any textbook on optics, see, \eg, \refref{salehteich}. 
However the results are broadly equivalent to a more general optical system.
Figure~\ref{f:4fsystem} shows a schematic setup. When applying the thin element approximation for the two lenses, 
one can relate any transverse electric (or magnetic) field component $\hat{u}$ in frequency space (see Appendix~\ref{a:fourier} for definitions) in the object plane ($y=y_o$) and the image plane ($y=y_i$) via
\beq\label{eq:4fsystem}
  \begin{split}
  & \hat{u}(x,y_i,z,\omega) = - e^{i4f\omega/c}\\
  & \quad \times \iint\limits_{k_{\perp}<\frac{\omega D}{2cf}} \bar{\hat{u}}(k_x,y_o,k_z,\omega)e^{-ik_xx-ik_zz}\mathrm{d}k_x\mathrm{d}k_z.
  \end{split} 
\eeq
Here, $D$ is the diameter of the aperture ($\hat{=}$ diameter of the mask in the Fourier plane), $f$ the focal length of the two identical lenses, 
and $ \bar{\hat{u}}(k_x,y_o,k_z,\omega)$ represents the field component in the object plane. 
In our configuration (propagation of the probe pulse in negative $y$ direction, $\hat{u}=\hat{E}_x,\hat{B}_z,\hat{E}_z$, or $\hat{B}_x$), 
we have $y_i=y_o-4f$, and the aperture blocks all Fourier components with transverse wave vector $k_{\perp}=\sqrt{k_x^2+k_z^2}>\frac{\omega D}{2cf}$. 
Equation~(\ref{eq:4fsystem}) is derived by assuming paraxial propagation in vacuum, 
and the two (perfect) lenses and the mask in the Fourier plane are treated in the thin element approximation. 

From our PIC simulations, we have access to a temporal snapshot of the fields $\vec{E}(\vec{r},t_S)$, $\vec{B}(\vec{r},t_S)$, 
at some time $t=t_S$ when the probe pulse has already passed through the interaction region and its center is at
$y_S$, \ie, at a distance $\Delta y = y_S-y_o$ from the object plane. 
So in general, at time $t=t_S$ the probe pulse has already passed the object plane, 
which usually lies somewhere inside the interaction region, but has not yet reached the first lens.
By symmetry of the $4f$-system, the electromagnetic fields at $y_S$ are the same as the fields 
at $y_S' = y_i+\Delta y$ at a distance $\Delta y$ from the image plane [up to a constant phase shift and image reversal, \cf \refeq{eq:4fsystem}].
The fields at the image plane could then be obtained by propagating the fields from point $y_S'$ to the image
plane or, by propagating the fields from position $y_S$ backwards to the object plane.
We choose to do the latter, thus circumventing the need to describe the full propagation of the pulse through
the imaging system. In our argument we ignored the effect of the aperture. However, since its role 
is to merely eliminate directions for which $k_{\perp}<\frac{\omega D}{2cf}$, it can be taken into account
independently of the position of the pulse, a fact that we will use in the following.

The (backwards) propagation from $y_S$ to the object plane is achieved by using the temporal snapshot of the electromagnetic field 
as initial datum to Maxwell's equations in vacuum. 
To this end, it is useful to remember that any solution to Maxwell's equations in vacuum $\vec{E}(\vec{r},t)$, $\vec{B}(\vec{r},t)$ can be decomposed into plane waves
\begin{align} 
 \vec{\mathcal{E}}_{\vec{k}}(\vec{r},t) &= \vec{\mathcal{E}}_{\vec{k},0}e^{i\vec{k}\cdot\vec{r}-i\omega t} \label{eq:evac} \\
 \vec{\mathcal{B}}_{\vec{k}}(\vec{r},t) &= \frac{\vec{k}\times\vec{\mathcal{E}}_{\vec{k},0}}{\omega}e^{i\vec{k}\cdot\vec{r}-i\omega t}, \label{eq:bvac}
\end{align}
where $\vec{k}$ is the wave vector, $\vec{\mathcal{E}}_{\vec{k},0}$ is the amplitude vector, $\vec{k}\cdot\vec{\mathcal{E}}_{\vec{k},0}=0$, 
and the vacuum dispersion relation dictates $\omega=kc$ with $k=|\vec{k}|=\sqrt{k_x^2+k_y^2+k_z^2}$.
For technical convenience, in the following we resort to complex field notation
\begin{align}
 \vec{E}(\vec{r},t)&=\vec{\mathcal{E}}(\vec{r},t)+\vec{\mathcal{E}}^*(\vec{r},t)=2\mathfrak{Re}\vec{\mathcal{E}}(\vec{r},t) \label{eq:ecomplex}\\
 \vec{B}(\vec{r},t)&=\vec{\mathcal{B}}(\vec{r},t)+\vec{\mathcal{B}}^*(\vec{r},t)=2\mathfrak{Re}\vec{\mathcal{B}}(\vec{r},t) \label{eq:bcomplex}\\
 \vec{\mathcal{E}}(\vec{r},t)&=\int \Theta(\omega) \hat{\vec{E}}(\vec{r},\omega)e^{-i \omega t} \mathrm{d} \omega \\
 \vec{\mathcal{B}}(\vec{r},t)&=\int \Theta(\omega) \hat{\vec{B}}(\vec{r},\omega)e^{-i \omega t} \mathrm{d} \omega,
\end{align}
and $\Theta(x)$ represents the usual Heaviside step function.
For given $\vec{\mathcal{E}}(\vec{r},t_S)$, $\vec{\mathcal{B}}(\vec{r},t_S)$ we can thus compute the fields at any time $t$ and at any position (in particular in the object plane) via
\begin{align}
 \vec{\mathcal{E}}(\vec{r},t) &= \iiint \bar{\vec{\mathcal{E}}}(\vec{k},t_S) e^{-i kc (t-t_S)}e^{i \vec{k}\cdot \vec{r}} \mathrm{d}^3k \label{eq:prope}\\
 \vec{\mathcal{B}}(\vec{r},t) &= \iiint \bar{\vec{\mathcal{B}}}(\vec{k},t_S) e^{-i kc (t-t_S)}e^{i \vec{k}\cdot \vec{r}} \mathrm{d}^3k,
 \label{eq:propb}
\end{align}
where $\bar{\vec{\mathcal{E}}}(\vec{k},t_S)$, $\bar{\vec{\mathcal{B}}}(\vec{k},t_S)$ denote the respective complex fields in spatial Fourier space.

\begin{figure}[ht!]
    \begin{center}
    \includegraphics[width=0.7\columnwidth, clip=true]{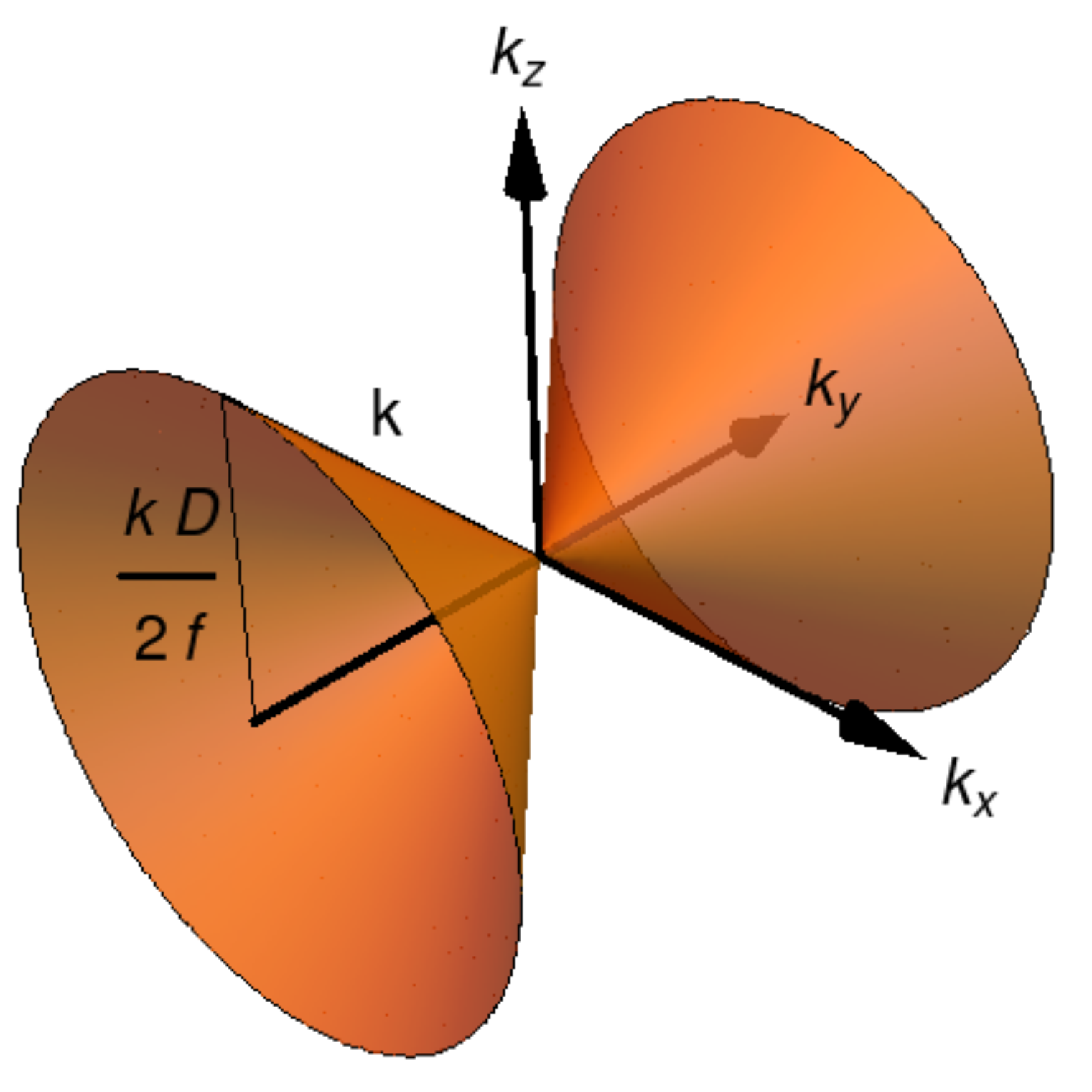} 
  \end{center}
 \caption{\label{f:probecone}The $4f$ system selects plane wave components in a cone defined by $k_{\perp}<\frac{k D}{2f}$ only. 
 Because the probe pulse propagates in negative $y$ direction, its complex electric field in Fourier space $\bar{\mathcal{E}}(\vec{k},t)$ is nonzero for $k_y<0$ only. 
 The same is true for the complex magnetic field $\bar{\mathcal{B}}(\vec{k},t)$.}
\end{figure}

Even though pump and probe pulse may overlap in position space at time $t_S$ of our snapshot, 
they are well separated in Fourier space due to their different propagation directions. 
Our $4f$ system selects plane wave components inside a cone defined by $k_{\perp}<\frac{k D}{2f}$, 
and therefore only the probe pulse gets imaged (see \reffig{f:probecone}).
Moreover, it is safe to assume that no light propagates in positive $y$ direction, i.e., in opposite direction to the probe pulse. 
This additional assumption allows us to write 
(see Appendix~\ref{a:complex} for the general case)
\begin{align}
\bar{\vec{\mathcal{E}}}(\vec{k},t_S) & =\Theta(-k_y)\bar{\vec{{E}}}(\vec{k},t_S) \\
\bar{\vec{\mathcal{B}}}(\vec{k},t_S) & =\Theta(-k_y)\bar{\vec{{B}}}(\vec{k},t_S). 
\end{align}
Hence, by using \refeq{eq:4fsystem}, the complex fields of the probe pulse in the image plane at any time $t$ are given by
\begin{align}
  \begin{split}
    & {\vec{\mathcal{E}}}(x,y_i,z,t) \\
  = & - \iiint\limits_{k_{\perp}<\frac{k D}{2f}} \bar{\vec{\mathcal{E}}}(\vec{k},t_S) e^{-i k[c (t-t_S)-4f]}e^{-ik_xx+ik_yy_o-ik_zz} \mathrm{d}^3k.
  \end{split} \\ \label{eq:image}
  \begin{split}
    & {\vec{\mathcal{B}}}(x,y_i,z,t) \\
  = & \iiint\limits_{k_{\perp}<\frac{k D}{2f}} \bar{\vec{\mathcal{B}}}(\vec{k},t_S) e^{-i k[c (t-t_S)-4f]}e^{-ik_xx+ik_yy_o-ik_zz} \mathrm{d}^3k.
  \end{split}
\end{align}
Finally, the shadowgram can be constructed by computing the $y$-component of the time integrated Poynting vector in the image plane
\beq\label{eq:Poynting}
\langle{S}_y\rangle(x,z) \propto \vec{e}_y \cdot \int_{-\infty}^{+\infty} \vec{E}(x,y_i,z,t) \times \vec{B}(x,y_i,z,t) \mathrm{d}t.
\eeq
We note that the propagation in time is transformed to propagation in space in
the practical implementation of our post-processing methodology, as explained in \refsect{s:method_imp}.

\subsection{\label{s:method_imp} Post-processing: Implementation}

We have seen in the previous section that in order to simulate shadowgrams, we have to 
\begin{enumerate}
 \item filter the relevant Fourier components of the snapshot $E(\vec{r},t_S)$, $B(\vec{r},t_S)$ 
		from the PIC simulation in order to extract the probe electromagnetic field components which will form the image,
 \item propagate the probe fields (in vacuum) to the object plane (resp.\ image plane), 
		and compute the time integrated Poynting vector to obtain the shadowgram.
\end{enumerate}
Let us now develop a post-processing algorithm doing exactly that. From our PIC code, 
once the probe pulse has passed the interaction region, 
we record the real field components $E_{\alpha}^{(i,j,m)}$, $B_{\alpha}^{(i,j,m)}$, 
$\alpha=x,z$ for each point $(i,j,k)$ on our computational grid, $i=0,\ldots, N_x-1$, $j=0,\ldots, N_y-1$,
$m=0,\ldots, N_z-1$. For ease of implementation
$N_x,\,N_y,\,N_z$ are assumed to be even (if not, we truncate the matrices
accordingly).
We cell-center the data which originally reside on a Yee grid. 
We perform a three dimensional spatial discrete Fourier transform of the field components in order to obtain 
$\bar{E}_{\alpha}^{(k_x,k_y,k_z)}$, $\bar{B}_{\alpha}^{(k_x,k_y,k_z)}$, where
$k_x\in\frac{2\pi}{N_x\, \Delta x}\,[-N_x/2,\, -N_x/2+1,\,\ldots, N_x/2-1]$, etc.
In the following, wavenumbers will be used as indices of matrices storing the spatial Fourier
transform of fields.

\begin{figure}[ht!]
  \begin{center}
    \includegraphics[width=\columnwidth, clip=true]{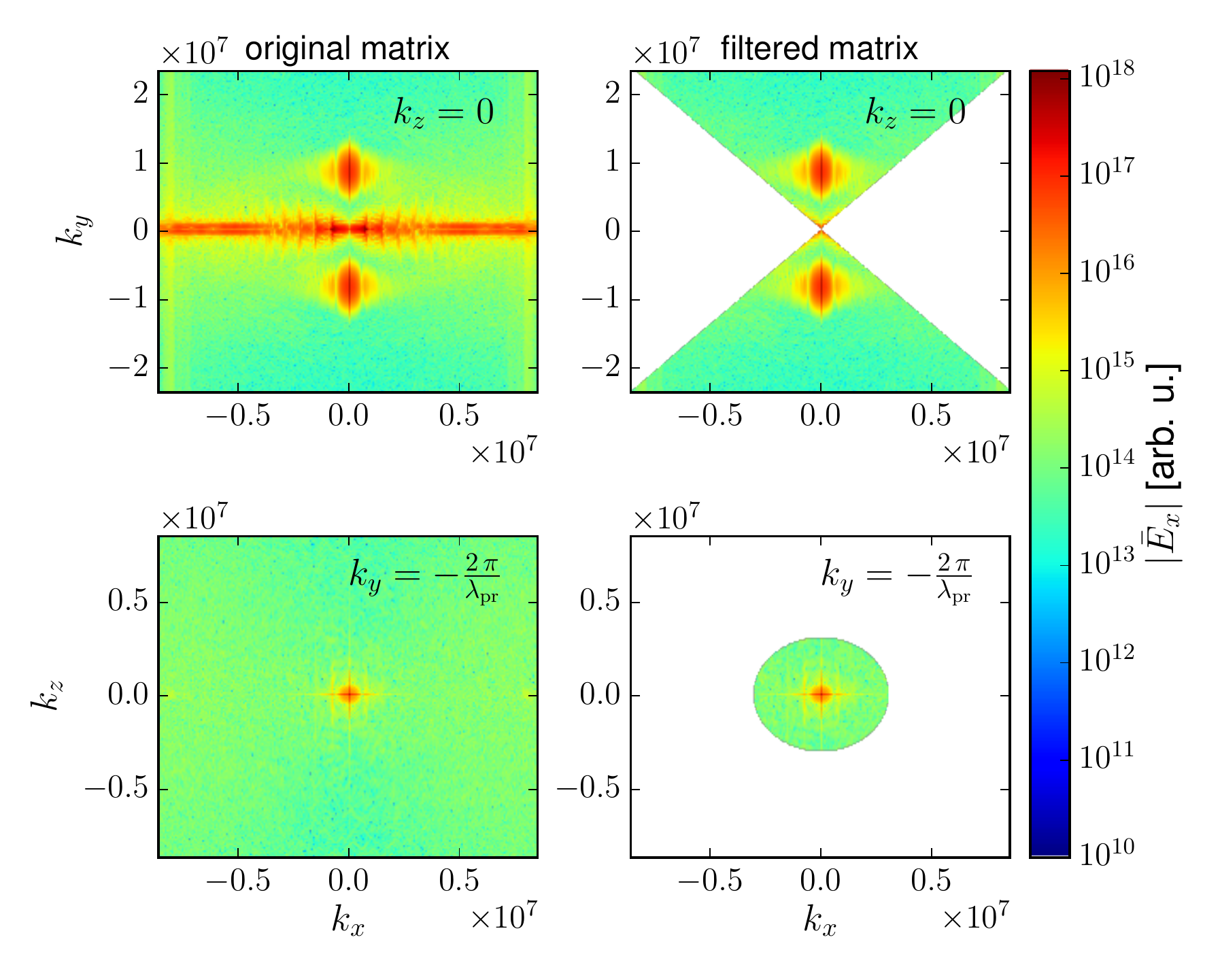} 
  \end{center}
  \caption{\label{f:na} Effect of taking into account the aperture in spatial Fourier space, visualized in the $k_x,k_y$ resp.\ $k_x,k_z$ plane. 
	  The value of the numerical aperture used here is $NA=0.26$.
	  Because we are using Fourier transforms of the real fields, both cones (positive and negative $k_y$) are present (\cf~\reffig{f:probecone}).
  }
\end{figure}

The first step is to set all matrix elements of matrices $\bar{E}_{\alpha}$, $\bar{B}_{\alpha}$
for which $k_{\perp}>kD/2f$ holds equal to zero. 
This models the effect of the aperture of the imaging system on our data,
and also eliminates the field of the pump pulse (see \reffig{f:na}). 
Note that because we computed the spatial Fourier transform of real fields, 
both cones (positive and negative $k_y$) feature nonzero Fourier amplitudes (\cf~\reffig{f:probecone}).
Because typical values of $D/f$ are small, most entries of the filtered matrices are zero. 
Thus, copying only the relevant matrix elements to new matrices $\bar{E}'_{\alpha}$, $\bar{B}'_{\alpha}$ of dimensions $N_x',\,N_y,\,N_z'$
significantly increases the efficiency of all further post-processing.

The second step is to propagate (in vacuum) the components 
of the probe electromagnetic field obtained above to the position of the object plane at $y=y_o$ 
and compute the time integrated Poynting vector, i.e., the shadowgram. 
According to \refeqs{eq:prope}{eq:propb}, propagation by distance $\Delta y = y_o-y$ is achieved
by multiplying components $\bar{E}_{\alpha}^{'(k_x,k_y,k_z)}$, $\bar{B}_{\alpha}^{'(k_x,k_y,k_z)}$
with
\beq
  P(\Delta y) = e^{-i\, \mathrm{sign}(k_y)\, k\,\Delta y}\,.
\eeq
Note that here we translated propagation time into propagation distance via $c\Delta t = \Delta y$, and the sign factor $\mathrm{sign}(k_y)$
in the exponent accounts for the fact that we deal with Fourier transforms of real fields, and not complex fields as in \refeqs{eq:prope}{eq:propb}.
Since in experiments the position of the object plane can vary from shot to shot by a few microns with respect to the position of the 
center of the wake, it is useful to compute shadowgrams for different positions of the object plane $y_o$
using the following approach:
We introduce a matrix $\langle{S}_y\rangle^{(i,j,m)}$ of the same dimensions as ${E}^{'}_\alpha$
to store time-integrated Poynting flux along $y$.
The fields are then propagated for $\Delta y\in[\Delta y^{(min)},\Delta y^{(min)}+\delta y,\ldots, \Delta y^{(max)}]$
where $\delta y\ll \lambdaPr$. Typically, one may use the spatial grid vector elements $y_s=y_j$ for $j<j_{max}$,
where $j_{max}$ should be chosen such that the probe pulse does not reach the boundaries of the numerical box.
Then, for each propagation distance $y_s$, we transform the propagated fields back to position space, 
form $\vec{e}_y \cdot \left( \vec{E}^{'} \times \vec{B}^{'} \right)$ and sum up all contributions in $\langle{S}_y\rangle$.
By construction, slices of the matrix $\langle{S}_y\rangle$ 
along the direction $j$ (corresponding to $y$) are proportional to
the time-integrated Poynting flux for an object plane at this position.
Of course, out of the $N_y$ possible object plane positions, we can only use those through
which the full length of the probe pulse was propagated during the above procedure.

A typical simulated shadowgram obtained by this procedure is shown in \reffig{f:probeTracking}(e)
for $y_o=0$. We can see that the scattered light from the pump which overwhelmed the image in 
\reffig{f:probeTracking}(d), has been eliminated through the introduction of an aperture
and subsequent suppression of large transverse wavevector components, and thus at the expense of spatial resolution. 
At the same time, as demonstrated in \refsect{s:example}, 
correcting for the object plane position has allowed to obtain an image that matches the typical
experimental shadowgrams\rf{saevert2015},  
where special care is taken to image approximately the center of the interaction region.

\subsection{\label{s:filt} Post-processing: Frequency Filtering}

In some cases it is desirable to account for frequency filters or polarizers in the experimental setup, 
or even the sensitivity curve of the CCD sensor itself.
Such filters can be simulated easily after step one of the above post-processing algorithm, i.e., in spatial Fourier space.
We just have to translate the frequency filter function given in $\omega$ to wavenumbers $k$ 
via the vacuum dispersion relation $k=\omega/c$, and multiply all field components. 
However, special care must be taken when
the frequency filter is such that it would lead to an effective increase of the duration of the probe pulse,
for example a narrow bandpass filter. In such case, we may have to increase the $y$ range of the computational box to accommodate the
filtered pulse, i.e., add some zero elements to the respective field matrices ${E}^{'}_\alpha$, $\bar{B}'_{\alpha}$.
Without these additional measures, we can have the situation that the filtered pulse
overlaps with itself by virtue of the periodic boundary conditions implied by the use of discrete Fourier transformations.

\section{\label{s:example} Comparison with experiments: An example}

The ability to track the probe propagation in our simulations, as in \reffig{f:probeTracking}(a-c), 
allows to interpret shadowgraphic images, such as those of \reffig{f:probeTracking}(e), and associate them
with transverse (to the direction of probe propagation) density gradients in the plasma. 
As we will show in this section, this is important in order to interpret experimental shadowgraphic images
and to deduce quantitative information from them.

\begin{figure}[ht!]
 \begin{center}
	(a)~\includegraphics[width=0.9\columnwidth]{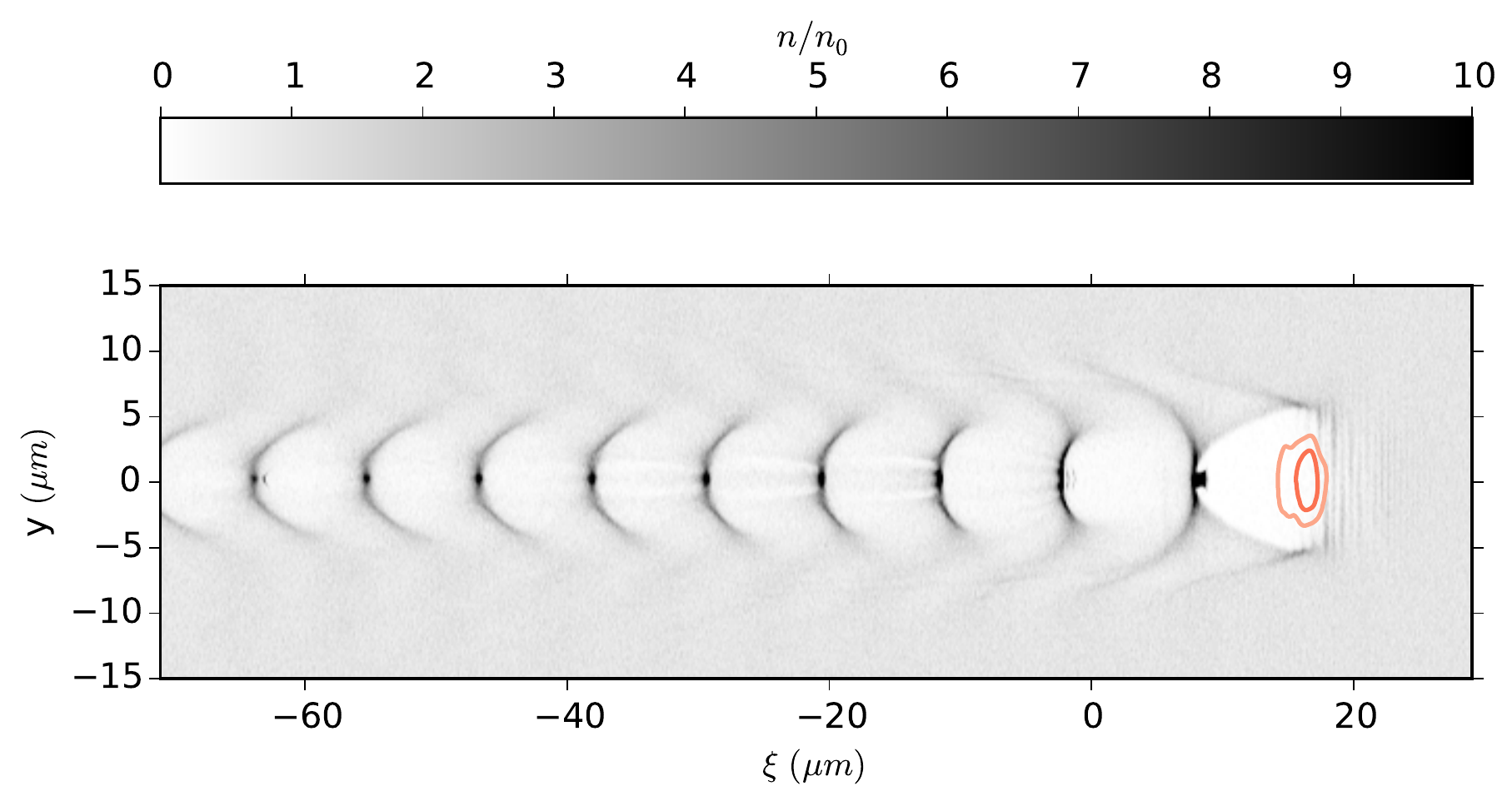}\\
	(b)~\includegraphics[width=0.9\columnwidth]{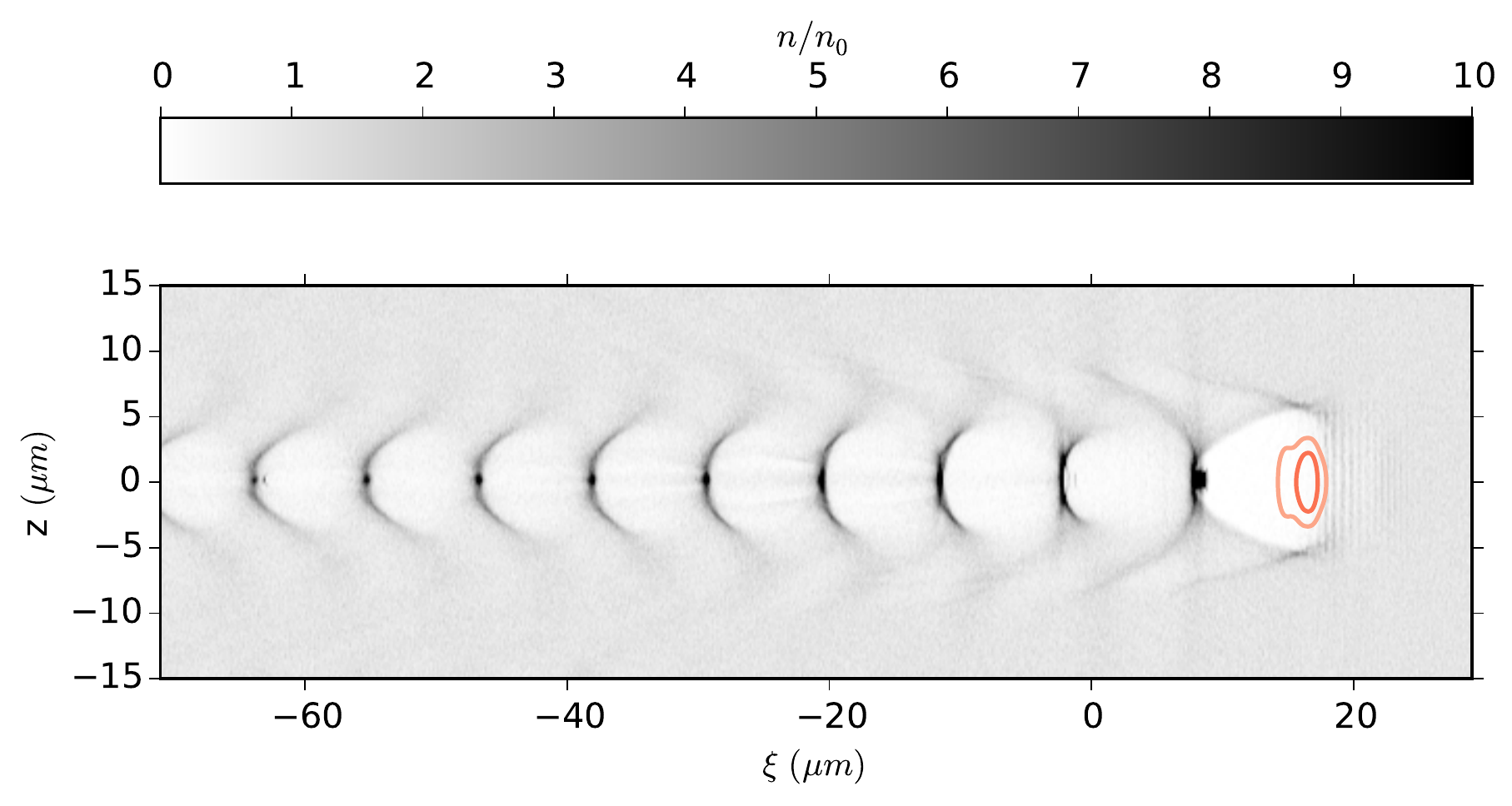}\\
	(c)~\includegraphics[width=0.9\columnwidth]{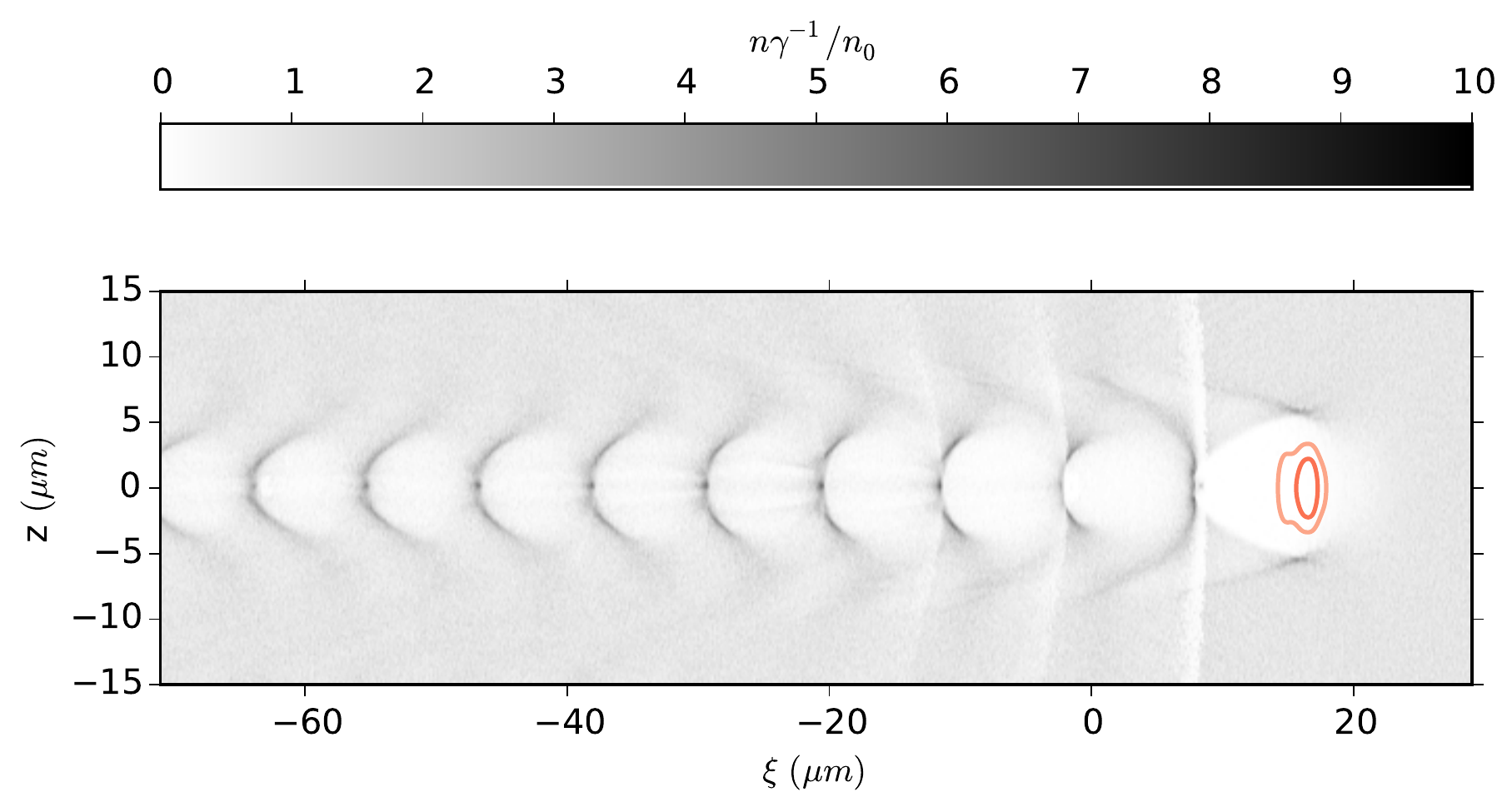}\\
	(d)~\includegraphics[width=0.9\columnwidth]{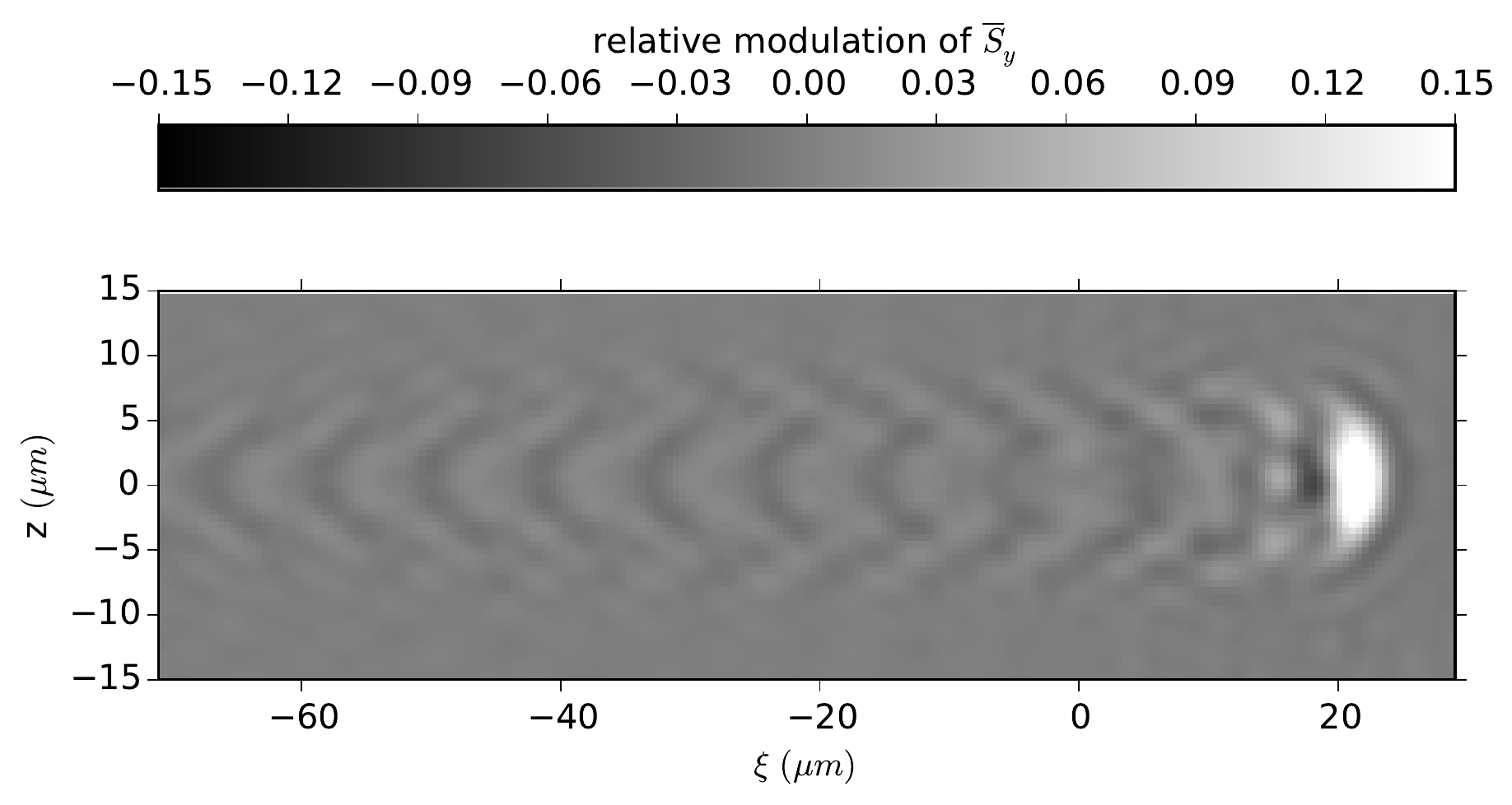}
 \end{center}
  \caption{\label{f:densShad} Snapshots of density distributions $n/n_0$ from the PIC simulation for (a) $z=0$ and (b) $y=0$
    at the time when the probe passes through $y=0$. We introduced the coordinate $\xi=x-ct$.
    (c) Map of the $y=0$ cross-section of $n/(\gamma\, n_0)$, (d) artificial shadowgram obtained with 
	probe pulse with duration $\tpr=12\,\fs$, central wavelength $\lambdaPr=750\,\mathrm{nm}$ 
	and object plane at $y_f=0$ (see \refsect{s:probe}). For the remainder of this paper we use the same
	pump-probe delay as in this figure and the same probe pulse characteristics unless otherwise specified.
	The color scale in all the following figures is the same as in panel (d).} 
\end{figure} 

In \reffig{f:densShad}(a-b) we plot cross-sections of the density distribution 
from a snapshot of our PIC simulation,
while the corresponding shadowgram is shown in \reffig{f:densShad}(d).
As \reffig{f:probeTracking} suggests, the main contribution is due to the strong density
gradients near the symmetry axis of the wake. Moreover, the strongest modulations 
originate at the front of the bubble, where the laser-pulse drives the wake. 

\begin{figure*}[ht!]
  \begin{center}
    \includegraphics[width=0.9\textwidth, clip=true]{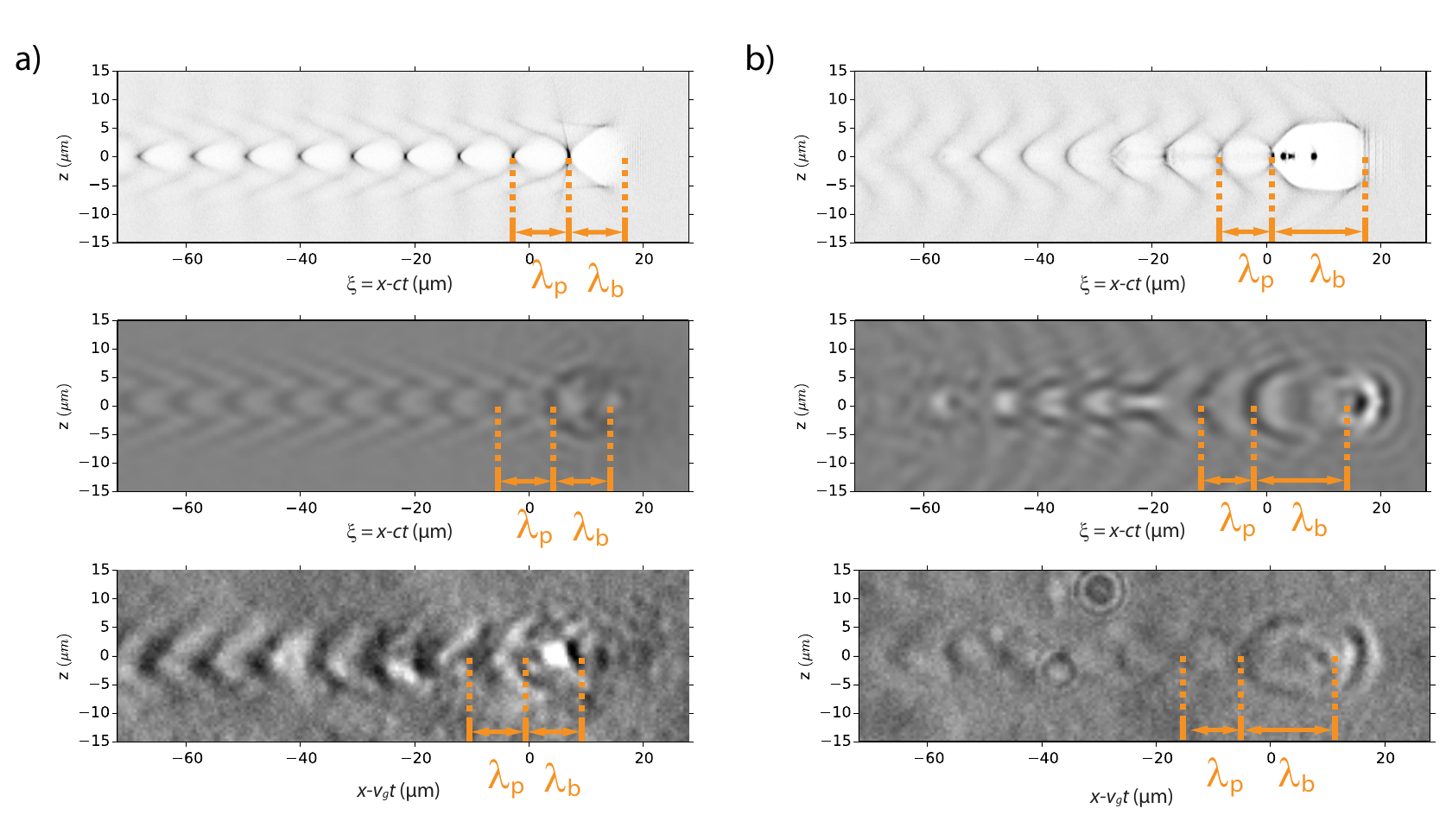} 
  \end{center}
  \caption{\label{f:bmeasure} Measuring the plasma wave length 
at different propagation distances a) $v_g t=527\,\mu m$ and b) $v_g t=1214\,\mu m$. Top: electron density, center: computed shadowgram, bottom: experimental shadowgram. 
  }
\end{figure*}

\begin{figure*}[ht!]
  \begin{center}
    \includegraphics[width=\textwidth, clip=true]{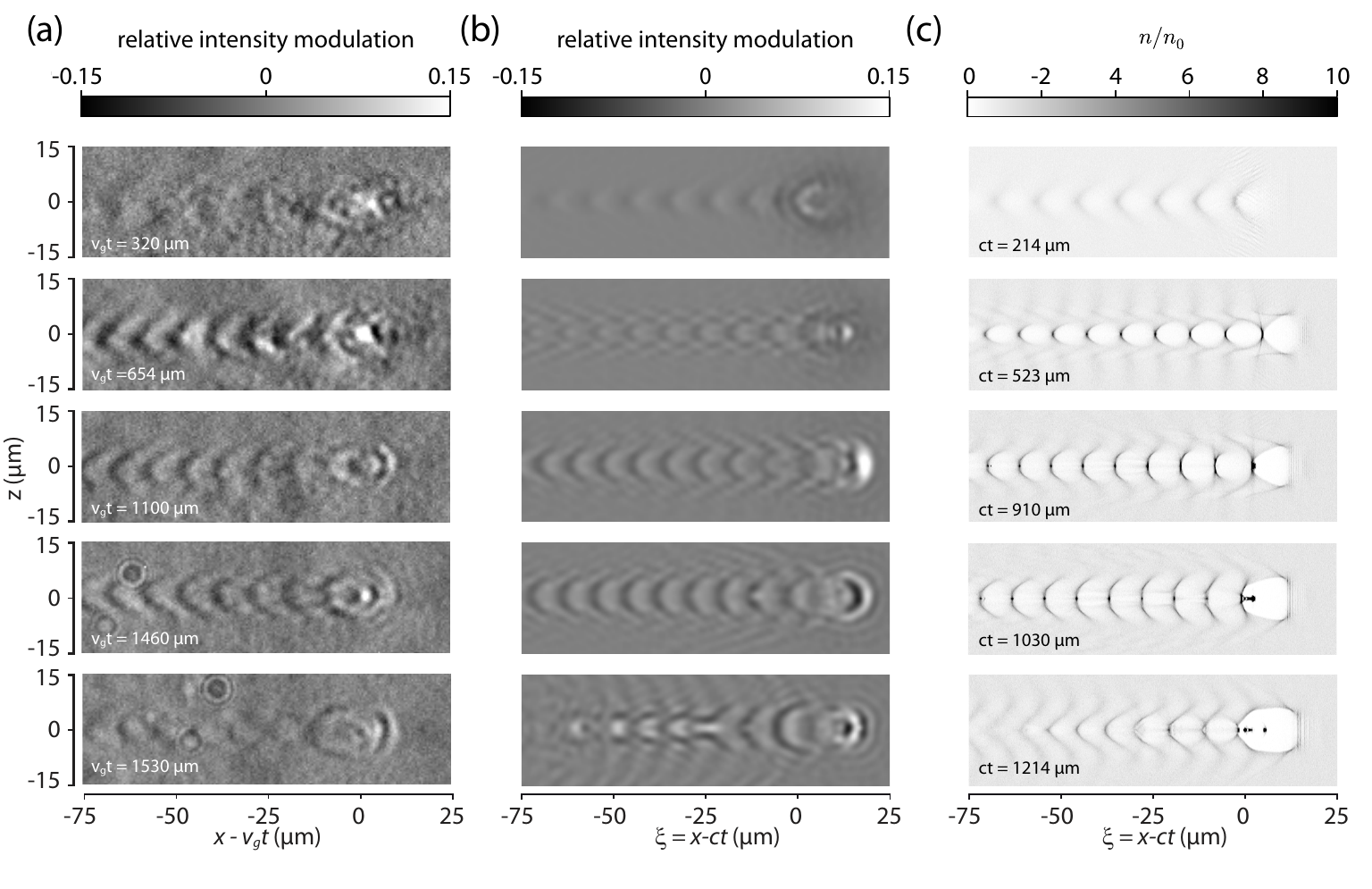} 
  \end{center}
  \caption{\label{f:picvsexp} (a) Experimental shadowgrams (b) computed shadowgrams, (c) density cross-sections from the simulation. 
  }
\end{figure*}

In \reffig{f:bmeasure} experimental shadowgrams (bottom row) 
taken with the few-cycle probing setup available at the JETI-laser facility at the Institute of Optics and Quantum Electronics in Jena, Germany, 
are shown. In the middle row we show computed shadowgrams obtained using the
same probe characteristics as in \reffig{f:densShad}, but also taking into account
the sensitivity versus wavelength of the CCD sensor used in the experiments.
Comparison of the experimental shadowgrams with the computed 
shadowgrams reveals that essential quantities like the period of the plasma 
oscillation, which can be easily retrieved in the electron density profile, are 
retained in the shadowgram. In \reffig{f:bmeasure} the 
length of the second oscillation period is defined by the distance between the 
dark regions in the intensity profile. 
In the PIC simulations the front of the bubble is readily identified as a peak in electron density. However
the experimental shadowgrams exhibit strong intensity modulations in this region which are not readily
recognisable as this density peak. The calculated shadowgrams also exhibit these modulations and hence
allow us to confirm that the centre of these modulations directly corresponds to the front of the bubble.
Crucially this then allows direct and quantitative measurements of the bubble shape and size from the
experimental shadowgrams.
Exemplary shots showing different 
stages during the evolution of the laser driven wake are plotted in \reffig{f:picvsexp}. In the first 
stage, in front of the position of the vacuum focus, only a small modulation in the 
plasma is visible. The phase fronts of the wave are slightly tilted due to a 
spatio-temporal asymmetry in the pump pulse. Later the pump pulse drives a 
plasma wave with a high amplitude. The transverse extent is of the order of the 
focal spot size and the phase fronts show the characteristic curvature
associated to nonlinear wakes\rf{bulanov1997}.
After wavebreaking and electron injection, 
the shape of the plasma wave, especially the back of the bubble, 
changes drastically. At the same time a strong modulation 
at the bubble front appears, which is also visible in the simulations. 

\begin{figure}[ht!]
  \begin{center}
    \includegraphics[width=\columnwidth, clip=true]{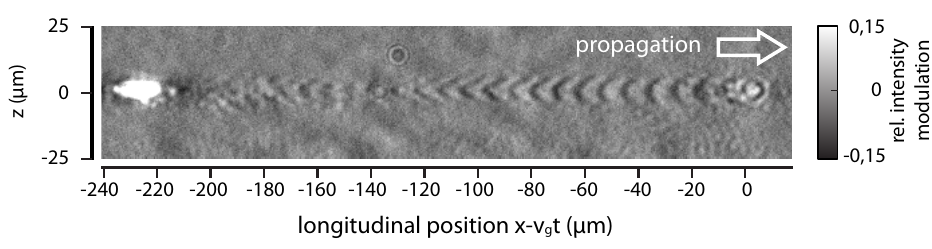} 
  \end{center}
  \caption{\label{f:wavebreaking} Experimental shadowgram of the laser wake field at $v_g t=1130\,\mu m$. The bright spot at $x-v_g t=-230\,\mu m$ is identified as wavebreaking radiation.
  }
\end{figure}

We observe that the largest
density gradient in \reffig{f:densShad}(a-b) is associated with the injected electron bunch inside the bubble.
However, this bunch is not visible in the shadowgrams. The reason is that the index of refraction
for a \emph{relativistic} plasma is:
$\eta=\left[1-\omega_{p}^2 /(\gamma\,\omega_{\rm{pr}}^2)\right]^{1/2}$, 
where $\omega_{\rm{pr}}$ is the probe frequency, $\gamma = \left(1+\overline{p}^2/m^2c^2\right)^{1/2}$ 
and $\overline{p}$ is the cycle-averaged electron fluid momentum.
The electrons of the injected bunch are therefore energetic
enough to be invisible to the probe pulse, as can be seen in \reffig{f:densShad}(d). This can be seen as another
facet of relativistic transparency\rf{akhiezer1956,kaw1970,cattani2000,siminos2012,palaniyappan2012}, 
where relativistic motion of electrons is induced by the wake field, while the modified index of refraction
is experienced by the probe pulse. We note that no signature of injected electrons
could be detected in the experimental shadowgrams of \refref{saevert2015}, 
even when injection occurred before the shadowgram was taken, as could be seen by the emission of wavebreaking radiation~\cite{thomas2007}.
Such an example experimental shadowgram is shown in \reffig{f:wavebreaking}. 
The simultaneously measured electron spectrum (not shown here) exhibits a broad energy range from 60\,MeV to 100\,MeV. 
At this time of the snapshot the laser pulse has propagated $1130\,\mu m$ into the gas jet. 
The bright spot on the left is attributed to wavebreaking radiation~\cite{thomas2007}. 
This is a direct signature for the onset of self-injection in the experiment. 
Relative to this point the laser pulse has traveled $230\,\mu m $ 
and electrons should be accelerated to an energy of $\approx 40\,$MeV ($\gamma=80$) 
assuming an effective electrical field of 160\,GV/m. 
{The typical charge measured by a spectrometer for the conditions of this experiment is $40\,\mathrm{pC}$.
Assuming that the bunch is ellipsoidal with volume $2\,\micron^3$ we find a number density $n_b=1.25\times10^{20}\,\mathrm{cm^{-3}}$.}
This implies no significant difference in the refractive index compared 
to the electron cavity surrounding the electron bunch. 
Thus the bunch remains invisible in the shadowgram. 

\section{\label{s:probeChar}Effect of probe pulse and imaging system parameters}

In this section we investigate changes in the simulated shadowgrams when parameters of the probe pulse
or the imaging systems are varied, which could be taken into account in the design of 
future experiments. As an important application we show that chirped pulses together with spectral filtering can,
in principle, be used to obtain time-dependent information from a single shot.

\subsection{Effect of object plane position}

\begin{figure}[ht!]
 \begin{center}
	(a)~\includegraphics[width=0.9\columnwidth]{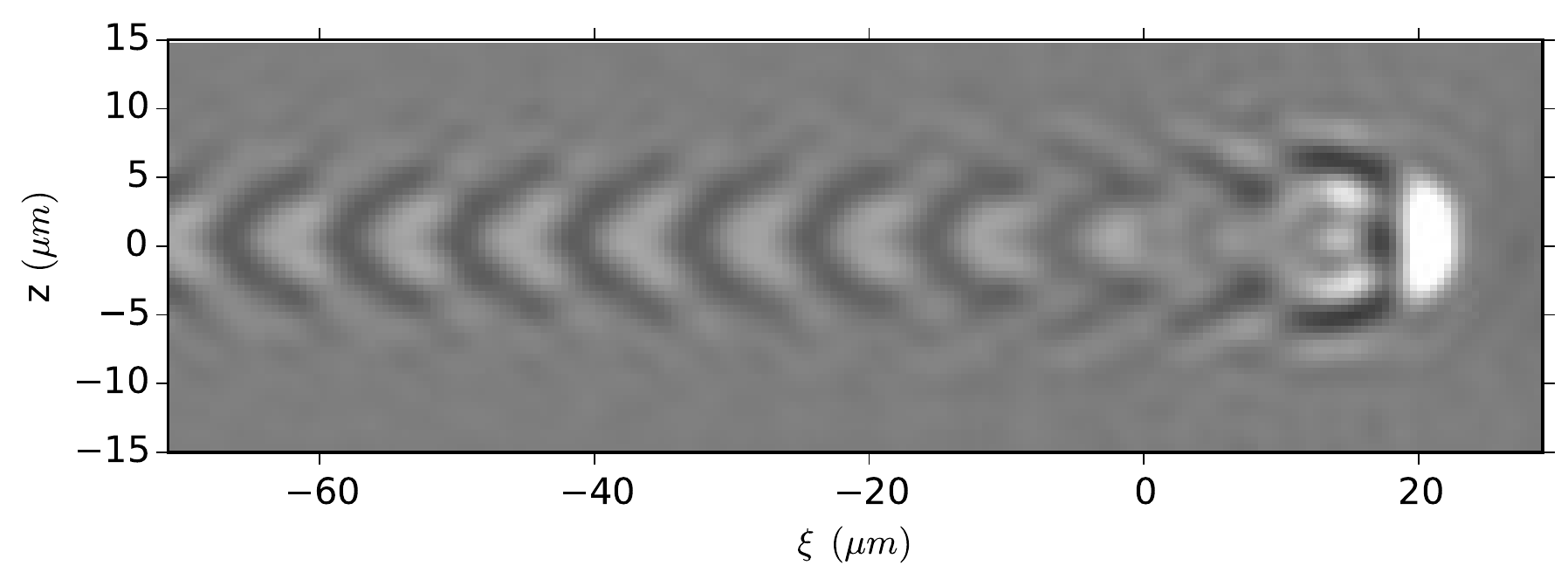}
	(b)~\includegraphics[width=0.9\columnwidth]{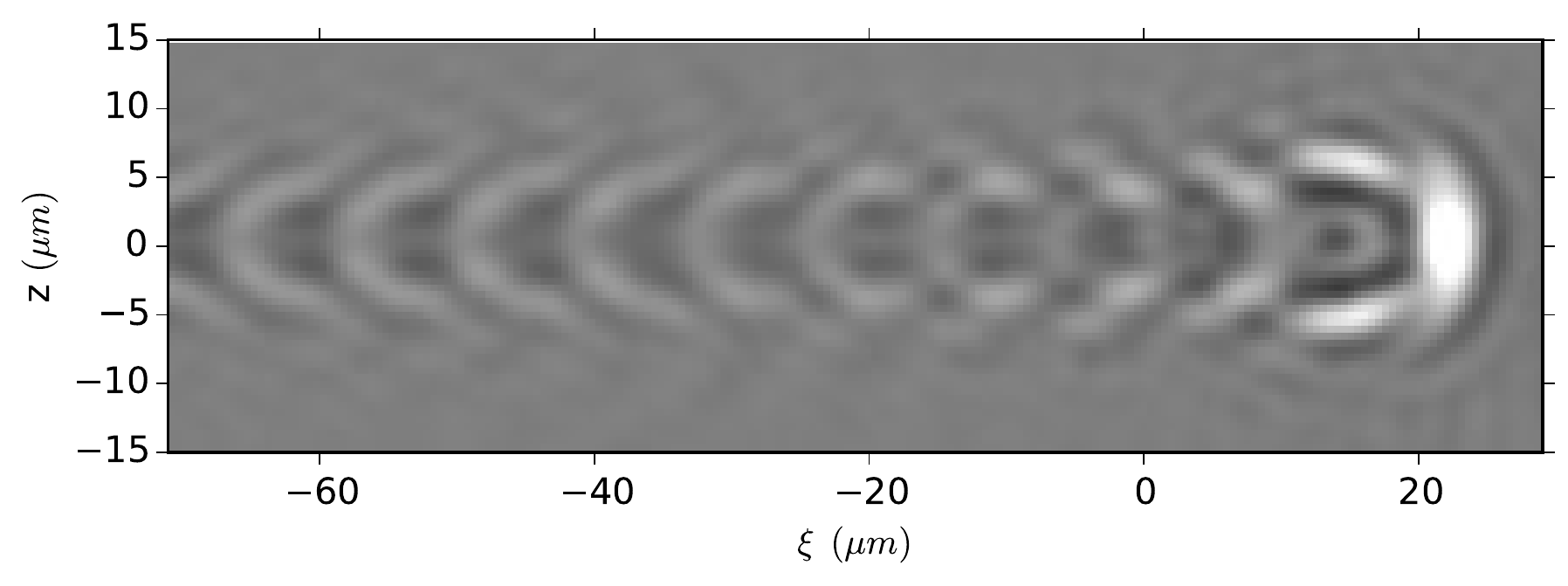}
 \end{center}
  \caption{\label{f:focus}Effect of varying the object plane position on the shadowgram (a) $y_o=-10\micron$, (b) $y_o=10\micron$. 
    All images have been obtained from the same pump-probe simulation used in \reffig{f:densShad}. 
    The object plane position has been varied in post-processing. To be compared to \reffig{f:densShad}(d) where $y_o=0$.
    }
\end{figure}

The post-processing methodology described in \refsect{s:post} allows us to adjust
the object plane position for the same probe propagation simulation. The effect
of object plane position on the produced shadowgram is shown in \reffig{f:focus},
where two shadowgrams have been reconstructed for different object plane positions $y_o=-10\micron$
and $y_o=10\micron$ from the same pump-probe simulation used in \reffig{f:densShad}. 
Changing object plane position not only brings different parts of the
wake into focus, but also results in enhanced blur due to diffraction compared to the case $y_o=0$
shown in \reffig{f:densShad}(d).

\subsection{Effect of probe pulse wavelength}

\begin{figure}[ht!]
 \begin{center}
   (a)~\includegraphics[width=0.9\columnwidth]{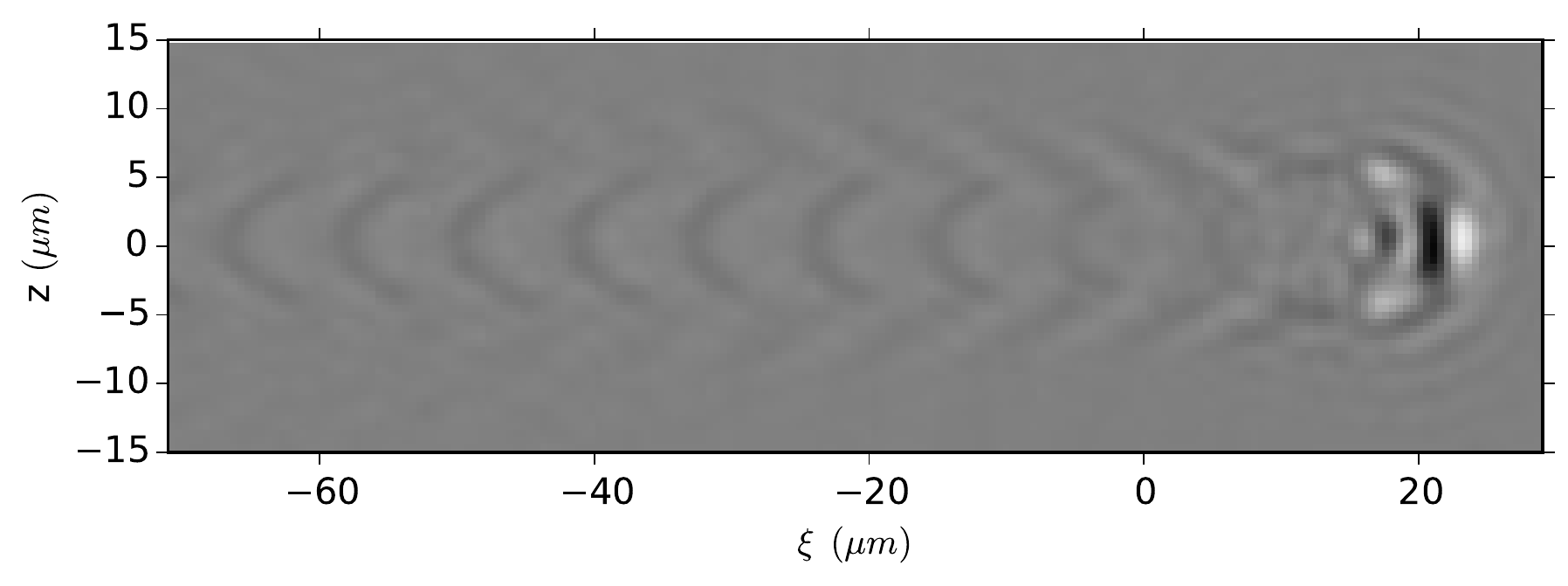}\\
   (b)~\includegraphics[width=0.9\columnwidth]{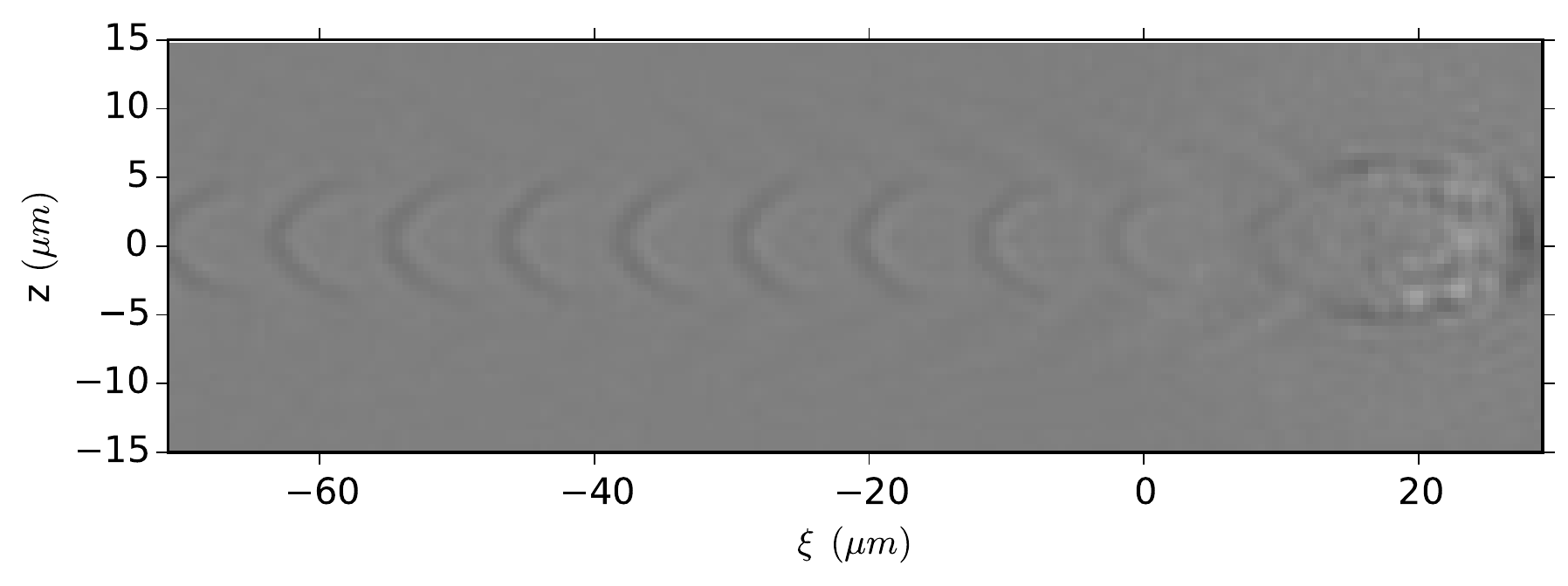}
 \end{center}
  \caption{\label{f:wavelength}Effect of varying probe pulse wavelength, while keeping its duration constant. 
  Both images have been obtained
   for the same pump-probe delay and the object plane position is fixed to $y_f=0$.
  (a) $\lambdaPr=0.6\,\micron$, (b) $\lambdaPr=0.45\,\micron$. To be compared to \reffig{f:densShad}(d) where
  $\lambdaPr=0.75\,\micron$. 
}
\end{figure}

Using a shorter wavelength probe pulse would be expected to reduce diffraction and improve resolution.
In \reffig{f:wavelength}(a) and (b) we see that this is indeed the case when $\lambdaPr$ is reduced to 
$0.6\micron$ and $0.45\micron$, respectively.
At the same time, since the refraction index of the plasma depends on the ratio of density to the critical
density (for the probe frequency) the variation in the index of refraction becomes
smaller for shorter wavelength, leading to somewhat lower contrast in the obtained images. 

\subsection{Effect of aperture}

\begin{figure}[ht!]
 \begin{center}
   \includegraphics[width=0.9\columnwidth]{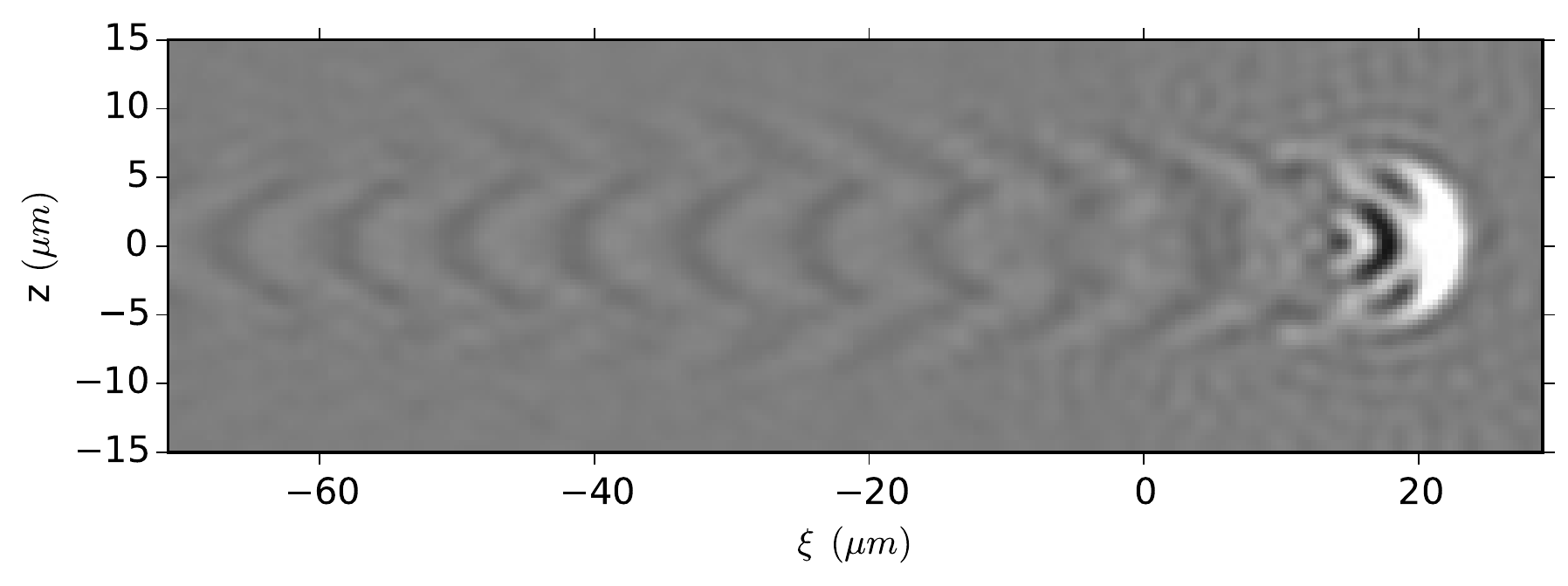}
 \end{center}
  \caption{\label{f:aperture} Same as \reffig{f:densShad}(d) but with $NA=0.35$. 
}
\end{figure}

The presence of an aperture in the imaging system prevents light scattered at large angles from 
reaching the image plane, while at the same time it limits the spatial resolution of the shadowgrams.
We illustrate the effect of varying the aperture acceptance angle in \reffig{f:aperture}, where
$NA=0.35$. Compared to \reffig{f:densShad}(d) for which $NA=0.26$ (matching the experimental one)
we see that in the present case the resolution is somewhat higher at the expense of increased
noise due to scattered light.

\subsection{Effect of probe pulse chirp}

Introducing a linear chirp in the probe pulse increases its duration (for a given spectral width)
and this is expected to result in a reduction in resolution of the shadowgrams due to motion blur. 
In our simulations a linear chirp is introduced through a quadratic phase in the probe pulse temporal profile,
\beq\label{eq:chirp}
  \phi = C (t-t_0)^2/w_t^2\,,
\eeq
where $C=\pm\left((\tpr/\tau_{BL})^2-1\right)^{1/2}$ is the chirp parameter, $\tau_{BL}$ is the bandwidth limited probe pulse duration, 
$t_0$ corresponds to the pulse maximum, and $w_t$ is connected to the intensity FWHM
duration of the probe pulse by $w_t=\tpr/\sqrt{2\ln2}$ and the positive (negative) sign of $C$ corresponds to a positive (negative) chirp.

\begin{figure}[ht!]
 \begin{center}
	(a)~\includegraphics[width=0.9\columnwidth]{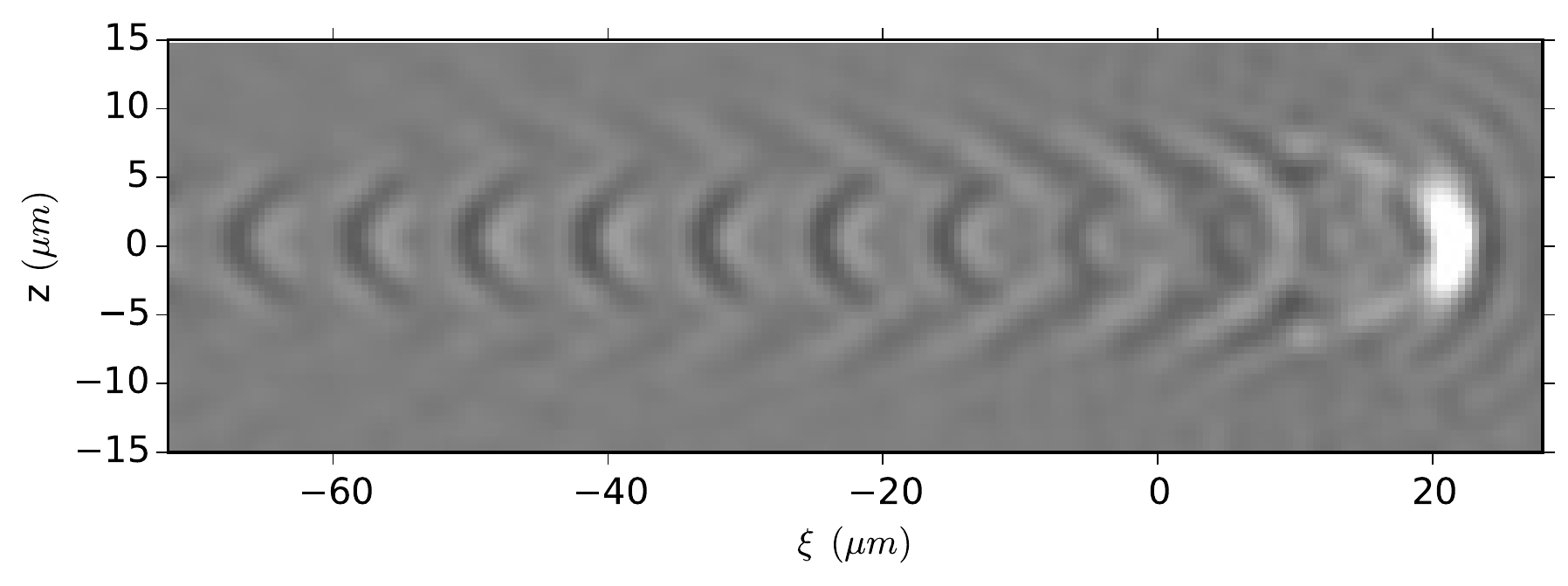}\\
	(b)~\includegraphics[width=0.9\columnwidth]{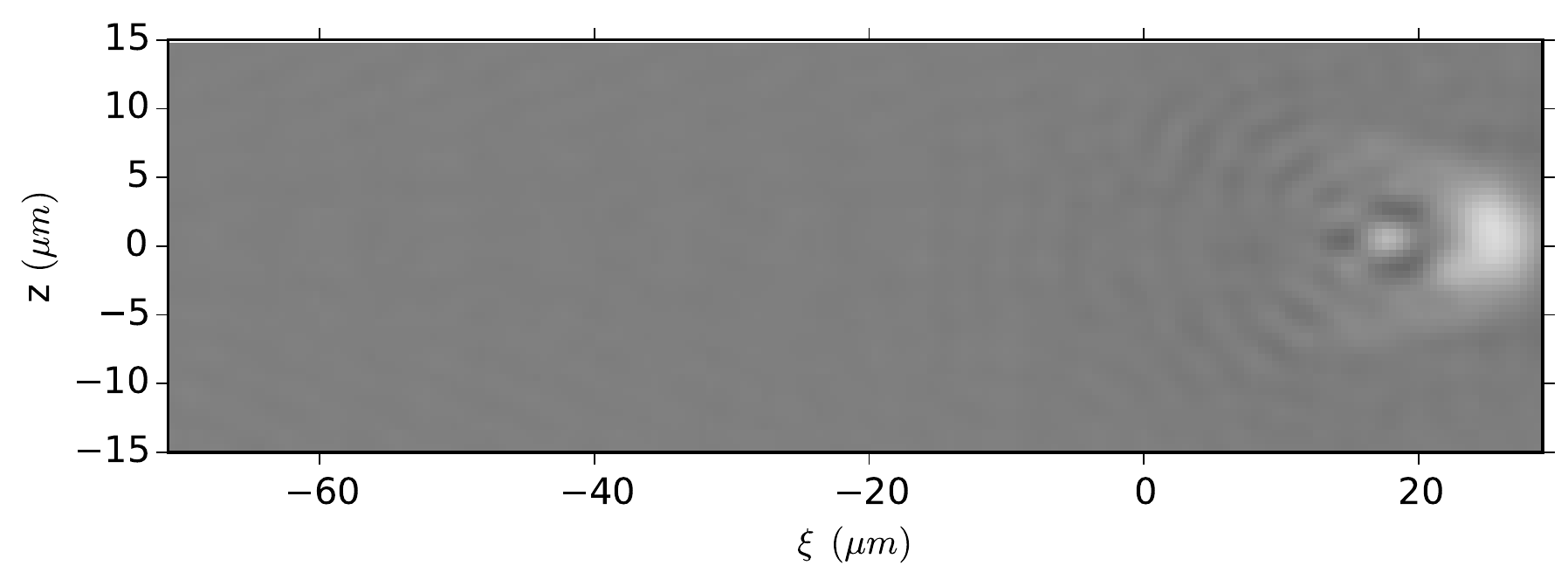}
 \end{center}
  \caption{\label{f:chirp}Effect of varying the probe pulse duration through the introduction of
  a linear chirp. (a) Fourier limited probe pulse
  duration $\tpr=4.4\,\fs$, (b) $\tpr=30\,\fs$. To be compared to \reffig{f:densShad}(d) where $\tpr=12\,\fs$.
  All images have been obtained for the same delay and the object plane position is fixed to $y_o=0$.
  }
\end{figure}

This is illustrated in \reffig{f:chirp}. We see that the probe pulse of \reffig{f:chirp}(a) with no chirp, \ie, having
the bandwidth-limited duration of $\tpr=\tau_{BL}=4.4\,\fs$ results in a shadowgram of higher resolution 
than the one of \reffig{f:densShad}(d), which was obtained with a pulse that had the same bandwidth but a duration 
$\tpr=12\,\fs$ (due to a negative chirp). Introducing a negative chirp such that $\tpr=30\,\fs$ results in the shadowgraphic image of \reffig{f:chirp}(b)
in which the structure of the wake is no longer visible. Since the length of such a pulse is $c\,\tpr\simeq9\,\micron$, which
is longer than $\lambda_p=8.1\,\micron$, the wake structure cannot be resolved in this case.

\subsection{Single shot movie generation through spectral filtering}

\begin{figure}[ht!]
 \begin{center}
	(a)~\includegraphics[width=0.9\columnwidth]{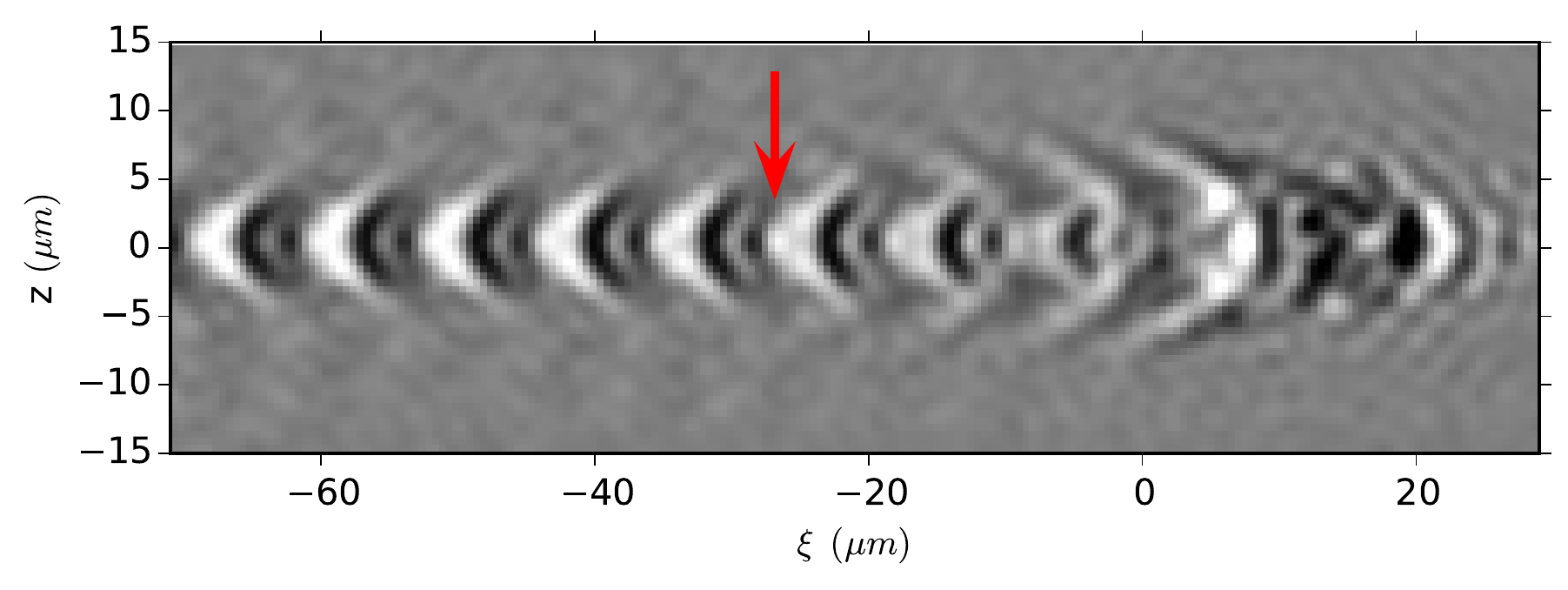}\\
 	(b)~\includegraphics[width=0.9\columnwidth]{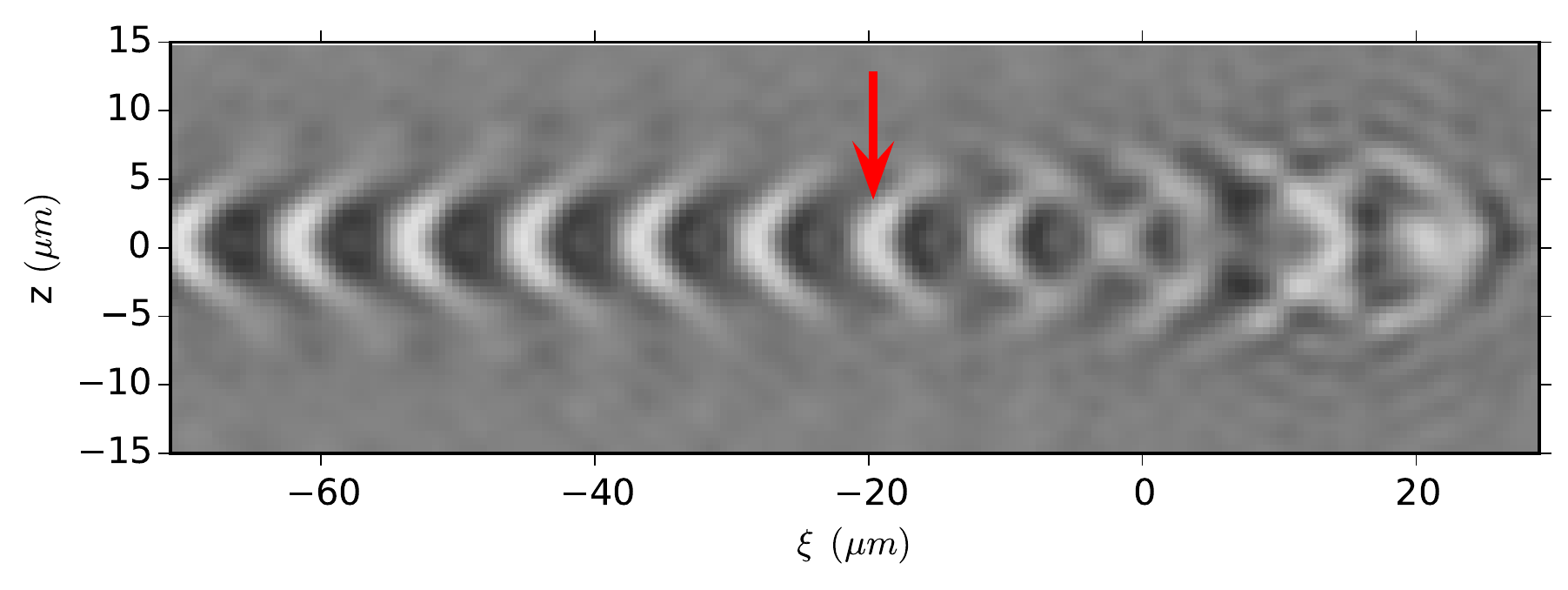}\\
	(c)~\includegraphics[width=0.9\columnwidth]{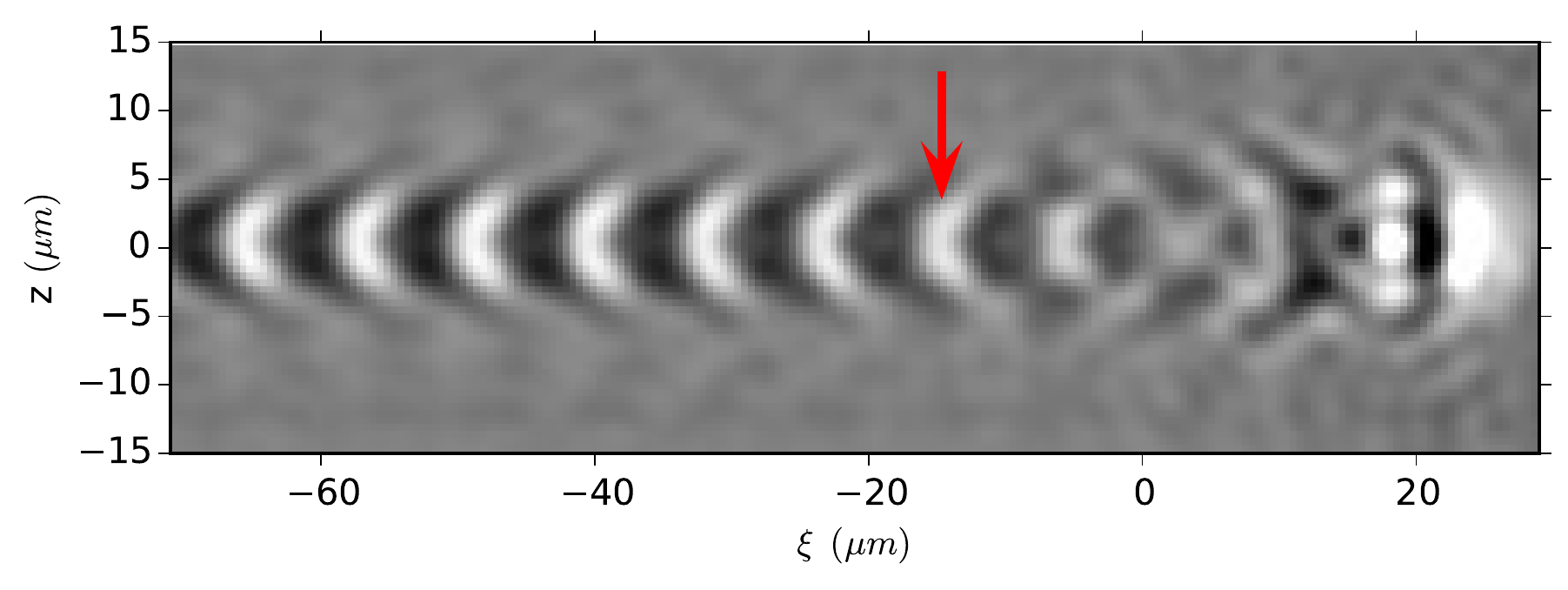}\\
	(d)~\includegraphics[width=0.9\columnwidth]{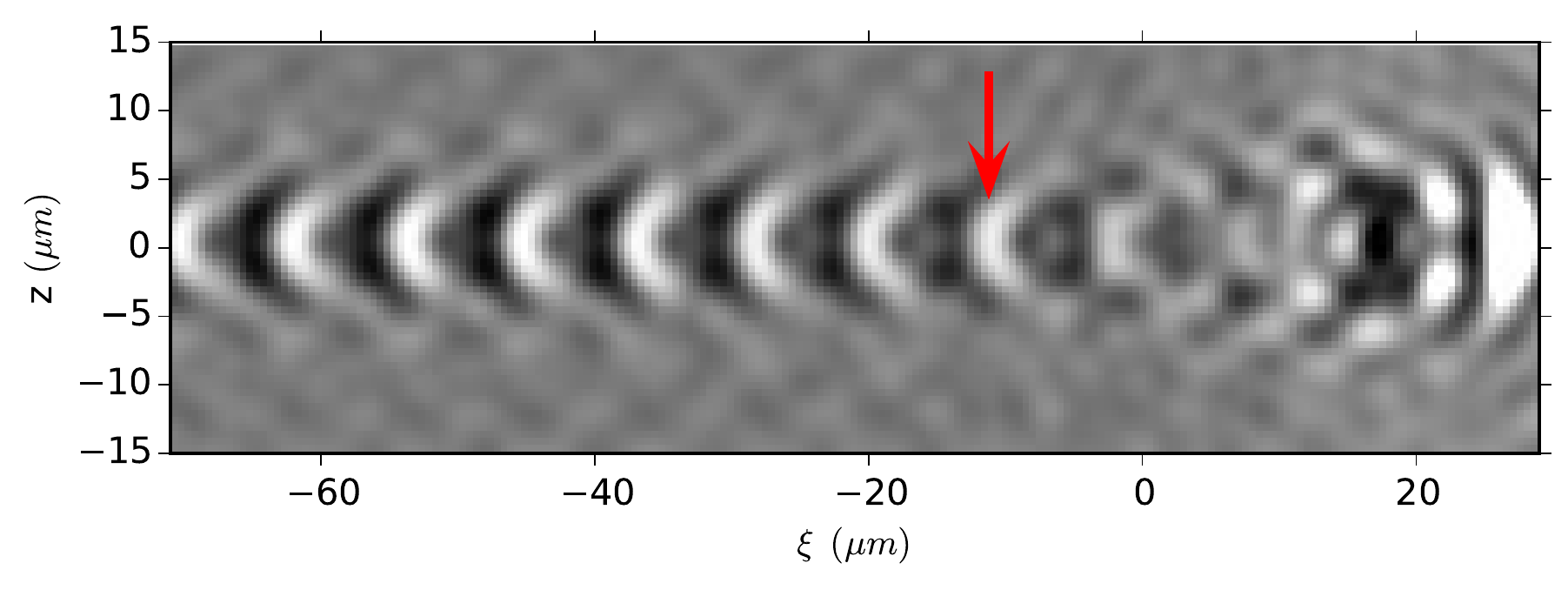}\\
 \end{center}
  \caption{\label{f:30fsFilt} Effect of introducing a bandpass filter after the $\tpr=30\,\fs$ chirped
  pulse of \reffig{f:chirp}(b) has propagated through the wake. 
  The filters allow transmission of wavelengths  
  (a) $0.6\pm0.01\,\micron$, (b) $0.7\pm0.01\,\micron$,
  (c) $0.8\pm0.01\,\micron$, (d) $0.9\pm0.01\,\micron$. 
  The red arrow indicates the position of the intensity maximum that we track in order to
  deduce the phase velocity of the wake.
   }
\end{figure}

The achievable temporal resolution for a linearly chirped pulse is given by the bandwidth limited pulse duration~\cite{polli2010},
and the information contained in the chirped $30\,\fs$ pulse of \reffig{f:chirp}(b) can be recovered by spectral filtering.
Introducing a step-like bandpass filter with central wavelength $\lambda_c$ in the range 0.6~to~0.9~\micron\
and width of $0.02\,\micron$ allows us to recover the shadowgrams of \reffig{f:30fsFilt}.
In these shadowgrams the signal to noise ratio is increased compared to the ones taken with shorter pulses,
since the fraction of the pulse energy contributing to the image formation is reduced.
Since the probe pulse has a negative chirp, shorter wavelengths interact with the wake earlier than longer wavelengths
and each frequency-filtered shadowgram captures a different snapshot of the motion of the wake in time.
This indicates that 1D time evolution of the wake could be obtained from a single pump-probe shot using a spectrometer after the interaction.
More complicated setups involving splitting and filtering the chirped probe pulse after the interaction could in principle allow 
the recovery of 2D time-dependent information. 
For example, we can estimate the wake phase velocity $v_p$ by measuring the position of a certain intensity maximum in
the different images of \reffig{f:30fsFilt}.
{In general one should take into account the plasma dispersion relation when analysing probe propagation,  
but here we will assume that the probe pulse chirp does not change significantly over the course
of propagation in the plasma and that we can use the vaccuum dispersion relation to connect frequency to wavelength.
This is justified for our simulations since the phase velocity of light in a tenuous plasma 
is given by $v_p\simeq c(1+n_0/n_c)$, where $n_c=\epsilon_0\,m_e\,\omega^2/e^2$ is the critical
density for frequency $\omega$. For the range of frequencies that we consider here, $5\times10^{-3}\lesssim n_0/n_c\lesssim10^{-2}$  
and the probe propagates in plasma for only a few tens of microns before reaching the interaction region in our simulations.}
Then, from \refeq{eq:chirp}, we find that the frequency of the probe pulse varies in time as
\beq
  \omega = 2 C (t-t_0)/w_t^2\,,
\eeq
and using the vacuum dispersion relation $\omega=c\,k$ to associate with wavelength $\lambda_c$ a time of interaction $t_c$,
we find
\beq\label{eq:t_lambda}
  t_c  = t_0 + \frac{\pi c w_t^2}{C}\left(\frac{1}{\lambda_c}-\frac{1}{\lambda_{pr}}\right)\,.
\eeq

Measuring the position of a certain intensity peak in the different images of \reffig{f:30fsFilt} and using
\refeq{eq:t_lambda} to estimate the corresponding time of interaction of the specific wavelength, \cf~\reftab{t:xtfilt},
allows us to estimate the phase velocity of the wake to be $v_{p}=2.8\times10^{8}\mathrm{m/s}$,
which is pretty close to the value we find by tracking the corresponding density maximum directly
in the PIC simulations, $v_{p}=2.9\times10^{8}\mathrm{m/s}$. 

\begin{table}[ht!]
  \begin{tabular}{c|c|r}
	$\lambda_1$ (\micron) 	& x (\micron)	& $\Delta t$ (\fs) \\ \hline
	0.6 					& -25.0        	& -30.2 \\
	0.7						& -19.6        	& -8.6 \\
	0.8 					& -14.7			& 7.5 \\
	0.9 					& -11.1			& 20.0
  \end{tabular}
 \caption{\label{t:xtfilt} Measurement of position $x$ of intensity maximum indicated by 
  a red arrow in \reffig{f:30fsFilt} for different central wavelength $\lambda_1$ of the applied filter
  and corresponding relative time of interaction $\Delta t=t_1-t_0$ computed using \refeq{eq:t_lambda}.
  }
\end{table} 

The length of the probe pulse in our computations is limited by the transverse extent of the computational domain to about $30\fs$.
During this time the length of the bubble does not change significantly.
However, longer pulses could capture the lengthening and change of shape of the bubble
in a single shot. 
We estimate that a pulse with transform limited duration $\tpr<20\,\fs$ chirped to $300\--600\,\fs$ would be sufficient to capture the lengthening 
of the bubble induced by pump pulse self-focusing and self-compression, according to the data presented in \refref{saevert2015}.
We note that a related technique which uses a train of short pulses of different wavelengths has recently been published~\cite{nakagawa2014}.

\section{\label{s:concl}Discussion and conclusions}

We presented a computational study of few cycle shadowgraphy of LWFA. Our study relies
on a combination of 3D PIC simulations of both pump and probe propagation in the plasma
and subsequent post-processing that takes into account the effect of a typical imaging
system. Both steps are crucial in order to analyze this technique.
We have compared the simulated shadowgrams with experimental ones and have shown that the former reproduce all
features found in the experiments. The ability to track probe propagation in the PIC simulations
allowed us to identify the signature of the front of the bubble in the shadowgrams. This in turn
facilitated the interpretation of the experimental shadowgrams and the measurement of important
quantities such as the length of the bubble~\rf{saevert2015}.

In order to facilitate the choice of probe pulse and imaging system parameters for future experiments, we studied
the effect of probe pulse wavelength and chirp, numerical aperture of the imaging optics and object plane position.
Moreover, we have shown that time-dependent information is retained in
a pulse with short transform limited duration (adequate to resolve the plasma wave)
that is linearly chirped to a longer duration. We envisage that if such a pulse is split after the interaction 
and filtered at different wavelengths, snapshots revealing the motion and evolution of the wake can be
obtained. 

In summary, we have proposed a method that allows us to construct synthetic ultrafast shadowgrams of plasma wakes 
by relating a near-field snapshot of the probe pulse to the integrated intensity at the image
plane of a typical imaging system. The technique presented here could be used to analyse similar 
experimental methods, \eg, sequentially timed femtophotography\rf{nakagawa2014}, or have applications in
other fields which involve the simulation of interaction of short probe pulses with matter, \eg, nanophotonics\rf{koenderink2015}.

\acknowledgments

For the computations, the supercomputer at the Max Planck Computing and Data Facility at Garching was used.
EPOCH was developed under UK EPSRC grants EP/G054940/1, EP/G055165/1 and EP/G056803/1. 
This study was supported by DFG (Grants No. TR18 B9, and No. KA 2869/2-1), 
BMBF (Contracts No. 05K10SJ2 and No. 03ZIK052), and European Regional Development Fund (EFRE). 
The collaboration was funded by LASERLAB-EUROPE (Grant No. 284464, EC’s Seventh Framework Programme).

\appendix

\section{\label{a:fourier}Definition of Fourier transforms}

 According to usual notations used in the context of optics, we define Fourier transforms with respect to space and time as follows: 
\begin{align}
\hat{f}(\omega)&=  \frac{1}{2\pi}\int f(t)e^{i\omega t}\mathrm{d}t\,, & {f}(t)&=\int \hat{f}(\omega)e^{-i\omega t}\mathrm{d}\omega\,, \\ 
\bar{f}(k_x)&=\frac{1}{2\pi}\int f(x)e^{-ik_x x}\mathrm{d}x\,, & {f}(x)&=\int \bar{f}(k_x)e^{ik_x x}\mathrm{d}k_x\,.
\end{align}

\section{\label{a:complex}How to compute complex electromagnetic fields from a temporal snapshot of real fields}

In the general case, without any prior knowledge on the propagation direction, one can compute the complex fields $\bar{\vec{\mathcal{E}}}(\vec{k},t_S),\bar{\vec{\mathcal{B}}}(\vec{k},t_S)$ from the real fields as follows: In spatial Fourier space we have according to \refeqs{eq:ecomplex}{eq:bcomplex} 
\begin{align}
\bar{\vec{E}}(\vec{k},t_S) & = \bar{\vec{\mathcal{E}}}(\vec{k},t_S) + \bar{\vec{\mathcal{E}}}(-\vec{k},t_S) \label{eq:ek} \\
\bar{\vec{B}}(\vec{k},t_S) & = \bar{\vec{\mathcal{B}}}(\vec{k},t_S) + \bar{\vec{\mathcal{B}}}(-\vec{k},t_S). \label{eq:bk}
\end{align}
Using Maxwell's equations in vacuum [cf. \refeqs{eq:evac}{eq:bvac}] 
\begin{equation}
\bar{\vec{\mathcal{B}}}(\vec{k},t_S) = \frac{\vec{k}\times\bar{\vec{\mathcal{E}}}(\vec{k},t_S)}{kc} \label{eq:be}
\end{equation}
we can rewrite \refeq{eq:bk} as
\begin{equation}
\bar{\vec{B}}(\vec{k},t_S) = \frac{\vec{k}\times\bar{\vec{\mathcal{E}}}(\vec{k},t_S) - \vec{k}\times \bar{\vec{\mathcal{E}}}(-\vec{k},t_S)}{kc}. \label{eq:bke}
\end{equation}
Thus, for given $\bar{\vec{E}}(\vec{k},t_S),\bar{\vec{B}}(\vec{k},t_S)$ we can solve Eqs.~(\ref{eq:ek}) and (\ref{eq:bke}) for $\bar{\vec{\mathcal{E}}}(\vec{k},t_S)$ and then compute $\bar{\vec{\mathcal{B}}}(\vec{k},t_S)$ from \refeq{eq:be}.

\end{document}